\documentclass[pra, twocolumn, showkeys, floatfix]{revtex4-2}
\usepackage{graphicx}
\usepackage{color}
\usepackage{amsmath, amsfonts, amssymb, bm}
\usepackage{tikz}
\usepackage{float}
\usepackage{subcaption}
\usetikzlibrary{decorations.pathmorphing}

\begin{document}
	\title{Coherently Enhanced Cherenkov Radiation by Highly Relativistic and Ultra-Compact Electron Beams}
	\author{S.~Kim}
	\author{C.~M\"uller}
	\author{A.~B.~Voitkiv}
	\affiliation{Institut f\"ur Theoretische Physik I, Heinrich-Heine-Universit\"at D\"usseldorf, Universit\"atsstra{\ss}e~1, 40225 D\"usseldorf, Germany}
	\date{\today}
	\begin{abstract}
Cherenkov radiation is emitted when charged particles penetrate dielectric media with velocity higher than the speed of light in the medium. Recent developments in plasma-based accelerator research enable creation of highly relativistic electron beams with femtosecond length and tens of micrometer scale radii, at energies of up to several GeV. Here, we consider Cherenkov radiation emitted by such extreme beams penetrating a gas target with refractive index very close to one. We demonstrate coherent enhancement in the emitted Cherenkov radiation up to the optical or even near-ultraviolet regime, which is due to the high beam energy and density. We also explore the dependence of the coherently enhanced Cherenkov radiation spectrum on the beam shape.
	\end{abstract}

	\maketitle
	
\section{Introduction}
Cherenkov radiation is emitted when charged particles move through a dielectric medium with velocity faster than the speed of light in the medium. Its experimental discovery by Cherenkov \cite{Cherenkov1,Cherenkov2} and theoretical explanation by Frank and Tamm \cite{FrankTammCherenkov,Cherenkov3} has been awarded with a Nobel prize in 1958 and the effect is known today for example as the "blue glow" in water surrounding nuclear reactor cores.\\
\indent Coherent enhancement can occur in the emission of radiation from a dense beam of particles. It arises when the contributions of the individual particles to the radiation interfere constructively. Already in 1954, Nodvick and Saxon \cite{NodvickSaxon} studied coherently enhanced synchroton radiation. Early considerations by Danos and Lahinsky \cite{Danos,Lahinsky} investigated Cherenkov diffraction radiation caused by a periodically modulated plane electron beam travelling near a dielectric medium to first theoretically demonstrate coherent enhancement in the emission of Cherenkov radiation. The importance of coherently enhanced Cherenkov radiation from cosmic particle showers was seen by Askaryan \cite{Askaryan1,Askaryan2} in the 1960s, leading to the coining of the Askaryan effect. The theory was then further specified for compact electron beams in the 1960s and 1970s \cite{Chaudhuri,Kukanov,Eastlund} and first experimental detections followed in the decades after, by various groups \cite{Buskirk,Ohkuma,Ciocci,Shibata}. The beams in these experiments had energies of tens of MeV, durations of a few picoseconds or longer and beam radii on the millimeter scale. Coherent enhancement was observed up to frequencies $\sim 2 \, \text{THz}$.\\
\indent In recent years, further experiments on coherently enhanced Cherenkov diffraction radiation by charged particles travelling near dielectric media were conducted. They include Cherenkov radiation from electron beams traveling through a dielectric waveguide, with sub-picosecond beam lengths, mm-scale beam radii and energies of few to few tens of MeV \cite{Cook,Pacey,Grigoryan,Veronese}; coherent enhancement was observed for frequencies $\lesssim 1\,\text{THz}$. Other experiments analyzed the spectrum of coherently enhanced Cherenkov diffraction radiation to gain information about the beam geometry, for electron beams of sub-picosecond lengths, sub-mm beam radius and energies of about $100\,\text{MeV}$, with coherent enhancement being observed for frequencies $\lesssim 2\,\text{THz}$ \cite{BeamDiag1,BeamDiag2}. Furthermore, tilting of the electron bunch to optimize the emission of coherent Cherenkov radiation was also explored for frequencies up to $\sim 1\,\text{THz}$, using electron beams of picosecond length, sub-mm beam radius and energy of few to few tens of MeV \cite{Tilt1,Tilt2}.\\
\indent Another recent focus of this research field has been coherent Cherenkov diffraction radiation from electron beams which consist of many periodic electron microbunches. This, in addition to the coherent enhancement caused by the coherent action of electrons within each microbunch, involves interference between contributions of different microbunches, which can lead to emission of radiation in form of nearly monochromatic lines \cite{FELCherenkov1,FELCherenkov2,Piazza}. This was also utilized to gain information about the microbunch length by analyzing the spectrum of emitted Cherenkov radiation \cite{BeamDiag1,FELCherenkov3}. The microbunch lengths used in these experiments were on the scale of picoseconds and the electron energies were in the range of tens of MeV. Coherent effects were observed up to frequencies $\sim 0.1 \, \text{THz}$.\\
\indent The limitation of coherent enhancement in past experiments to relatively low frequencies in the millimeter-wave or microwave regime is due to the rather large size of the utilized beams, both in transverse and longitudinal direction.\\
\indent Experimental capabilities in accelerator physics have improved immensely, especially recent developments in plasma-based acceleration make it possible to generate highly relativistic and very compact electron beams of unprecedented densities with energies up to $\sim 10$ GeV \cite{PlasmaBeams1,PlasmaBeams2,PlasmaBeams3,PlasmaBeams4}. For example, very recent experiments used the plasma photocathode method aiming to achieve electron beams of few-femtosecond lengths with currents of $\gtrsim 50 \, \text{kA}$ and energies of several GeV \cite{PlasmaPhotocathode1,PlasmaPhotocathode2}. As we will show, for such highly relativistic beams we can enhance coherence in the transverse direction to the beam velocity by choosing a dielectric target material with refractive index very close to $1$.\\
\indent In this paper, we consider coherent Cherenkov radiation by highly relativistic electron beams with femtosecond-scale length and micrometer-scale radius penetrating a helium gas and show that coherent enhancement can extend up to optical or even near-ultraviolet frequencies. To this end, we also provide scaling laws for the maximum coherently enhanced frequency as a function of the beam parameters and medium properties and calculate energy spectra as well as photon number spectra of the emitted Cherenkov radiation. We also explore the dependence of coherent enhancement on the electron beam shape by comparing emission spectra for a Gaussian, super-Gaussian \cite{SuperGauss} and cylindric constant-density beam \cite{Gelfer} and demonstrate that the coherent enhancement can have varying structures and also extend to different maximum frequencies.\\
\indent The paper is organized as follows: In section \ref{SectionTheory}, we give a brief overview of the theory of coherent Cherenkov radiation and derive expressions for the spectra of emitted photons. We then proceed with the calculation of the coherent enhancement for a Gaussian, super-Gaussian and cylindric constant-density beam. In section \ref{SectionResults}, we show the calculated spectra and demonstrate the qualitative differences in coherence between the different beam models. Lastly we give some concluding remarks in section \ref{SectionConclusion} and provide some mathematical details in the appendix.\\
\indent Gaussian units are used to express the theoretical formulas and atomic units (a.u.) in the presentation of our numerical results.

\section{Theoretical Considerations}\label{SectionTheory}
In the following, we briefly discuss the theory of coherent Cherenkov radiation \cite{FrankTammCherenkov,Eastlund} to lay a foundation for our subsequent discussion. Mathematical details, for the sake of brevity, are provided in the appendix.\\
\indent We consider a beam of $N_e$ electrons, each moving along the $z$ axis with velocity $v$, described by the trajectories

\begin{equation}
{\bf r}_\ell(t) = {\bf r}_{\perp,{\ell}} + v(t-t_{\ell}) \, {\bf e}_z, \quad \ell = 1,2,...,N_e\,.
\label{Trajectories}
\end{equation}

\noindent The corresponding charge and current densities are

\begin{equation}
\begin{aligned}
\rho({\bf r}, t) &= \sum_{{\ell}=1}^{N_e} \rho_{\ell}({\bf r},t) \quad,\quad {\bf j}({\bf r}, t) = v \, \rho({\bf r},t) \, {\bf e}_z\,,
\end{aligned}
\label{ChargeCurrentDensity}
\end{equation}

\noindent with $\rho_{\ell}({\bf r}, t) = e \delta({\bf r}_{\perp} - {\bf r}_{\perp,{\ell}}) \, \delta(z-v(t-t_{\ell}))$, where $e$ is the electron charge.\\
\indent We assume an infinitely extended, isotropic medium with particle density $n_0$ and presume that the number of particles $\lambdabar^3 n_0 = \left(\frac{c}{\omega}\right)^3 n_0 \gg 1$ for all frequencies considered, such that we can assume the medium to be homogenous. At the end of section \ref{SectionTheory}, we will specify the actual conditions on the medium size such that it can be considered "infinite". Let us additionally assume no magnetization in the medium, i.e. $\mu = 1$ and a dielectric permittivity $\varepsilon(\omega)$, such that Maxwell's equations written in terms of the potentials are

\begin{equation}
\begin{aligned}
\Delta\phi - \frac{1}{c^2} \frac{\partial^2\hat{\varepsilon}\phi}{\partial t^2} &= -\frac{4\pi}{\hat{\varepsilon}}\rho \;\; , \;\; \Delta{\bf A} - \frac{1}{c^2} \frac{\partial^2\hat{\varepsilon}{\bf A}}{\partial t^2} = -\frac{4\pi}{c}{\bf j}\,.
\end{aligned}
\label{Maxwell}
\end{equation}

\noindent Here, $\hat{\varepsilon}$ is an operator which is evaluated by application to the different frequency components contained in the potentials.\\
\indent Performing Fourier transformation (see appendix), we can express the Fourier transforms of the fields as

\begin{equation}
\begin{aligned}
\tilde{{\bf E}}({\bf k}, \omega) &= 2ie\frac{\delta(\omega - vk_z)}{k^2 - \frac{\varepsilon\omega^2}{c^2}}\Biggl[\frac{\omega v}{c^2}\,{\bf e}_z - \frac{\bf k}{\varepsilon}\Biggr]\\
&\quad \times \sum_{{\ell}=1}^{N_e} e^{-i{\bf k}_{\perp}\cdot{\bf r}_{\perp,{\ell}}} \, e^{i\omega t_{\ell}}\,,\\
\tilde{{\bf B}}({\bf k}, \omega) &= \frac{\varepsilon v}{c} \Bigl[-\tilde{E}_y \, {\bf e}_x + \tilde{E}_x \, {\bf e}_y \Bigr]\,,
\end{aligned}
\label{FourierFields}
\end{equation}

\noindent where $\varepsilon(\omega)$ is now a complex scalar. The fields are products of the single particle fields multiplied by the sum over phase factors corresponding to the $N_e$ particles in the beam. If the phases are small,  the sum is $\sim N_e$ and thus the spectra would be enhanced by a factor $\sim N_e^2$, i.e. full coherent enhancement. The "smallness" of the phases depends on the size of the beam compared to $k_{\perp}^{-1}$ and $k_{\parallel}^{-1} = \frac{v}{\omega}$, where $k_{\perp}$ is, at this point, still undetermined. Since the magnetic field can be obtained from the electric field, we only proceed with expressions for the latter.\\
\indent We apply inverse Fourier transformation only with respect to the ${\bf k}$-space to obtain

\begin{equation}
\begin{aligned}
&{\bf E}_{\omega}({\bf r}, \omega) = \frac{2ie}{(2\pi)^{3/2}v}e^{i\frac{\omega z}{v}}\sum_{{\ell}=1}^{N_e} e^{i\omega t_{\ell}}\\
&\hspace{0.5cm}\times\int_{\mathbb{R}^2} d^2{\bf k}_{\perp} \, \frac{\frac{\omega v}{c^2}(1 - \frac{c^2}{\varepsilon v^2})\,{\bf e}_z - \frac{{\bf k}_{\perp}}{\varepsilon}}{k_{\perp}^2 - k_0^2} \, e^{i{\bf k}_{\perp} \cdot \Delta {\bf r}_{\perp,{\ell}}}\,,
\end{aligned}
\label{E_omega}
\end{equation}

\noindent where we have introduced

\begin{equation}
\begin{aligned}
k_0 := \frac{\omega}{v}\sqrt{\frac{\varepsilon v^2}{c^2} - 1}\;,\;\Delta {\bf r}_{\perp,{\ell}} := {\bf r}_{\perp} - {\bf r}_{\perp,{\ell}}\,.
\end{aligned}
\label{k0_Delta_r_perp_Definition}
\end{equation}

\noindent Then we can write for the longitudinal field component

\begin{equation}
\begin{aligned}
E_{\omega,z} &= -\frac{2\pi^2e}{(2\pi)^{2/3}c^2} \, \omega e^{i\frac{\omega z}{v}} \Bigl(1 - \frac{c^2}{\varepsilon v^2}\Bigr)\\
&\quad\times\sum_{{\ell}=1}^{N_e} H_0^{(1)}(k_0 \, \Delta r_{\perp,{\ell}}) \, e^{i\omega t_{\ell}}\,,
\end{aligned}
\label{E_omega_z_Result}
\end{equation}

\noindent where $H_\nu^{(1)}$ denotes the Hankel function of the first kind. For the transverse component, we obtain (see appendix)

\begin{equation}
\begin{aligned}
{\bf E}_{\omega,\perp} &= \frac{2\pi^2iev}{(2\pi)^{3/2}c^2} \, k_0 \, e^{i\frac{\omega z}{v}} \frac{c^2}{\varepsilon v^2}\\
&\quad \times \sum_{{\ell}=1}^{N_e} H_1^{(1)}(k_0 \, \Delta r_{\perp,{\ell}}) e^{i\omega t_{\ell}} \frac{\Delta{\bf r}_{\perp,{\ell}}}{\Delta r_{\perp,{\ell}}}\,.
\end{aligned}
\label{E_omega_perp_Result}
\end{equation}

\indent Since we are interested in radiation, we consider the far-field zone and additionally assume that it is far away from the beam itself, i.e. $\vert k_0\vert r_{\perp} \gg 1$ and $r_{\perp} \gg r_{\perp,\ell}$. We can then use the asymptotic expansion of the Hankel functions

\begin{align*}
H_{\nu}^{(1)}(z) \approx \sqrt{\frac{2}{\pi z}} \, e^{i(z-\frac{1}{2}\nu\pi - \frac{1}{4}\pi)}\,.
\end{align*}

\noindent Moreover, we approximate

\begin{align*}
\Delta r_{\perp,\ell} \approx r_{\perp} - \frac{{\bf r}_{\perp} \cdot {\bf r}_{\perp, \ell}}{r_{\perp}}\,.
\end{align*}

\noindent Then it follows from eqs. \eqref{E_omega_z_Result} and \eqref{E_omega_perp_Result}:

\begin{equation}
\begin{aligned}
E_{\omega,z} &\approx -\frac{e}{c^2} \, \omega e^{i\frac{\omega z}{v}} e^{-i\frac{\pi}{4}} \frac{1 - \frac{c^2}{v^2\varepsilon}}{\sqrt{k_0}} \frac{e^{ik_0 r_{\perp}}}{\sqrt{r_{\perp}}} \, S\,,\\
{\bf E}_{\omega, \perp} &\approx \frac{iev}{c^2} \sqrt{k_0} \, e^{i\frac{\omega z}{v}} \frac{c^2}{\varepsilon v^2} \frac{e^{ik_0 r_{\perp}}}{\sqrt{r_{\perp}}}\, S \, {\bf e}_{\perp}\,,
\end{aligned}
\label{E_omega_approx_result}
\end{equation}

\noindent where ${\bf e}_{\perp} := \frac{{\bf r}_{\perp}}{r_{\perp}}$ and

\begin{equation}
S := \sum_{\ell =1}^{N_e} \, e^{-i{\bf k}_{\perp,0} \cdot {\bf r}_{\perp,\ell}} e^{i\omega t_{\ell}} \;, \quad {\bf k}_{\perp,0} := k_0 \, {\bf e}_{\perp}\,.
\label{S_Definition}
\end{equation}

\indent The far-field expressions in eq. \eqref{E_omega_approx_result} depend on the complex wave vector

\begin{equation}
\boldsymbol{\kappa} := \left(k_0\,{\bf e}_{\perp},\frac{\omega}{v}\right) \equiv (\boldsymbol{\kappa}_{\perp}, \kappa_z)\,,
\label{WaveVectorCherenkov}
\end{equation}

\noindent where the real part of $\boldsymbol{\kappa}$ corresponds to oscillation and the imaginary part to exponential decay of the electromagnetic field in the medium.\\
\indent Up to this point we have made no assumptions other than the initial assumptions about the medium itself. The distinction between a fast-decaying near-field and Cherenkov radiation lies in the value of $k_0$:\\
\indent The Cherenkov condition for far-fields seen from the definition of $k_0$ and the exponential function $e^{ik_0 r_{\perp}}$ in eq. \eqref{E_omega_approx_result} is

\begin{equation}
\frac{c}{v \sqrt{\text{Re}(\varepsilon)}} = \frac{c}{v\sqrt{n^2 - \varkappa^2}} < 1\,,
\label{CherenkovCondition}
\end{equation}

\noindent where $n$ is the refractive index and $\varkappa$ is the extinction coefficient, such that $\varepsilon = (n+i\varkappa)^2$. Neglecting absorption, one obtains the well-known Cherenkov condition $\frac{c}{nv} < 1$. If eq. \eqref{CherenkovCondition} is fulfilled and if we are far from absorption resonances in the medium, $k_0$ will be practically real such that the exponential $e^{ik_0 r_{\perp}}$ is an oscillating function with a weak damping factor. Conversely, if eq. \eqref{CherenkovCondition} is not fulfilled, the fields decay exponentially in $r_{\perp}$. Therefore, only frequencies $\omega$ for which eq. \eqref{CherenkovCondition} is fulfilled correspond to Cherenkov radiation.\\
\indent Now that we obtained far-field expressions for the Fourier-transformed fields, we continue with the energy spectrum of emitted radiation corresponding to the flow of energy through the area $A$:

\begin{equation}
\frac{d\mathcal{E}}{d\omega}\Biggl\vert_A = \frac{c}{4\pi} \int_A d{\bf A} \, 2 \, \text{Re}\bigl[{\bf E}_{\omega}(\omega) \times {\bf B}_{\omega}^*(\omega)\bigr]\,.
\label{dE_domega_general}
\end{equation}

\begin{subsection}{Radiation Through the Mantle of a Cylindrical Volume}\label{SubsectionMantleSpectrum}
We first consider the spectrum corresponding to the flow of energy through the mantle $M_c$ of a cylinder of radius $R_c$ with its length $L_c$ along the z-axis. We assume that the electron beam is cylindrically symmetric around the z-axis such that we can define an effective beam radius $R_b$ and effective beam length $L_b$.\\
\indent To avoid near-field contributions, we choose an asymptotic mantle radius $R_c$ in the far-field region which is additionally large compared to the beam dimensions,

\begin{align*}
R_c \gg R_b \;,\; \vert k_0 \, R_c\vert \gg 1\,.
\end{align*}

\noindent Then we can employ the far-field expressions \eqref{E_omega_approx_result} for the fields. Using eq. \eqref{FourierFields} and defining

\begin{equation}
\begin{aligned}
\xi_1(\omega) &:= \text{Re}\left[ \frac{(1 - \frac{c^2}{\varepsilon v^2})\sqrt{\frac{\varepsilon v^2}{c^2} - 1}}{\Bigl\vert \sqrt{\frac{\varepsilon v^2}{c^2} - 1} \Bigr\vert} \right]\,,
\end{aligned}
\label{dE_domega_mantle_help}
\end{equation}

\noindent we obtain (see appendix)

\begin{equation}
\begin{aligned}
\frac{d\mathcal{E}}{d\omega}\Biggl\vert_{M_c} &\approx \frac{L_c e^2}{2\pi c^2} \, \omega \, \xi_1(\omega) e^{-2\text{Im}(k_0) R_c} \int_0^{2\pi} d\varphi \, \left\vert S \right\vert^2\,.
\end{aligned}
\label{dE_domega_mantle_result1}
\end{equation}

\indent The average of the coherence factor $\vert S\vert^2$ over the azimuthal angle $\varphi$ is indicative of the amount of coherent enhancement in the spectrum, with the maximum being full coherence if the electron beam is sufficiently dense, yielding a factor $N_e^2$. In section \ref{SubsectionCoherenceFactor_AngularRelations} we will discuss the form of $\vert S\vert^2$ in more detail.\\
\indent Since we assume a cylindrically symmetric beam, $\vert S\vert^2$ will be independent of $\varphi$. Applying this to eq. \eqref{dE_domega_mantle_result1} and writing the spectrum doubly differential in frequency and $z$, we have

\begin{equation}
\frac{d^2\mathcal{E}}{d\omega dz}\Biggl\vert_{M_c} \approx \frac{e^2}{c^2} \, \omega \, \text{Re}[\xi_1(\omega)] e^{-2\text{Im}(k_0) R_c} \left\vert S \right\vert^2\,.
\label{dE_domega_dz_mantle_result2}
\end{equation}

\noindent Fully neglecting absorption, i.e. setting $\text{Im}(\varepsilon) = 0$, we arrive at the expression

\begin{equation}
\frac{d^2\mathcal{E}}{d\omega dz}\Biggl\vert_{M_c} \approx \frac{e^2}{c^2} \, \omega \left(1 - \frac{c^2}{\varepsilon v^2}\right) \, \left\vert S \right\vert^2\,.
\label{dE_domega_dz_mantle_result3}
\end{equation}

\noindent This is the well-known Frank-Tamm formula \cite{FrankTammCherenkov} for the Cherenkov spectrum, multiplied by the coherence factor $\left\vert S \right\vert^2$.

\end{subsection}

\begin{subsection}{Radiation Through the Top of a Cylindrical Volume}\label{SubsectionTopSpectrum}
For highly relativistic electron beams, we can choose a medium with a refractive index very close to $1$ while still fulfilling the Cherenkov condition \eqref{CherenkovCondition}. This, as we will see in section \ref{SubsectionCoherenceFactor_AngularRelations}, can strongly increase the effect of coherence.

\begin{figure}[H]
\centering
\includegraphics[width=5.5cm]{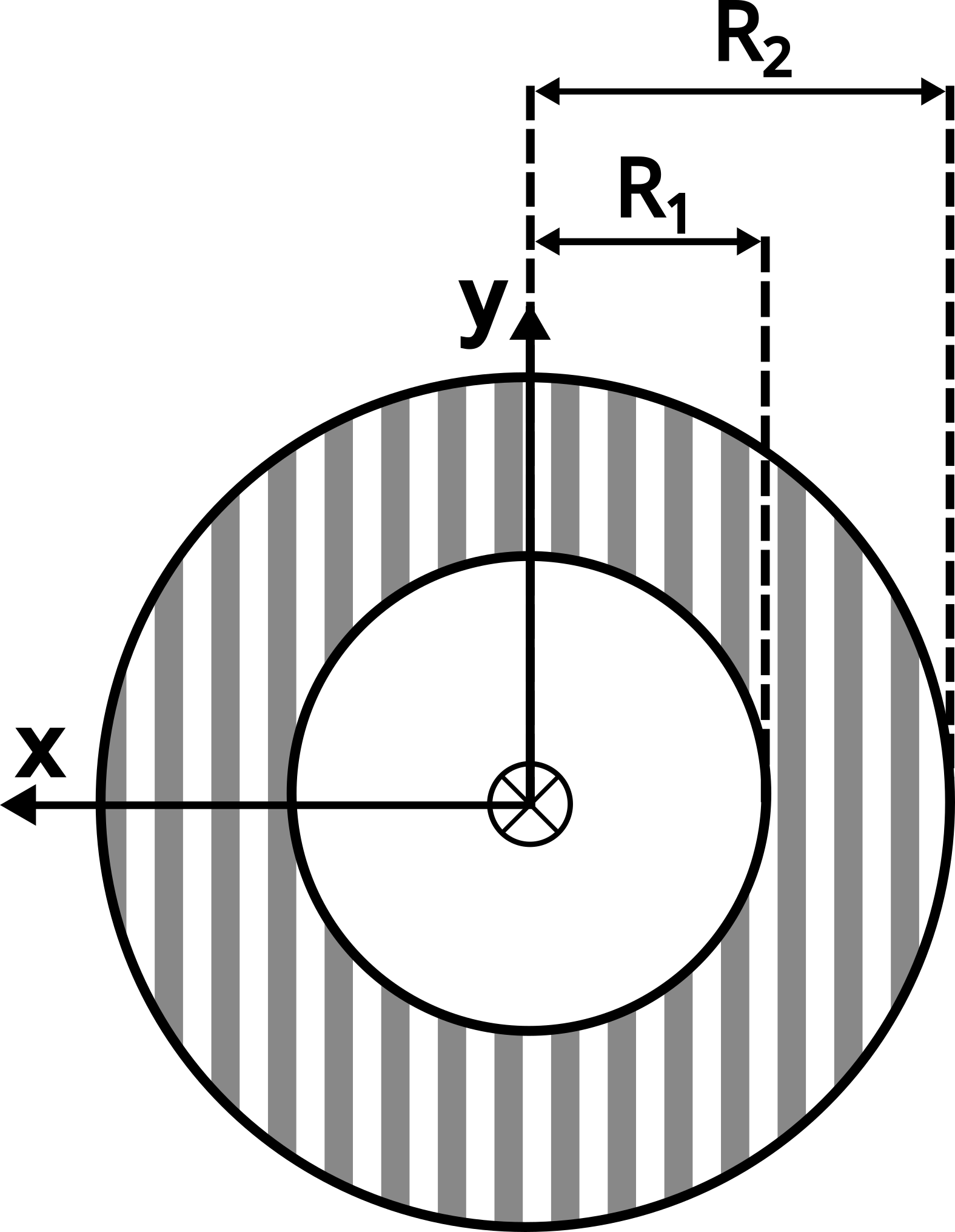}
\caption{Sketch of the ring-shaped detector with inner radius $R_1$ and outer radius $R_2$. Electron beam moving along the z-axis going into the plane in this sketch.}
\label{Figure1}
\end{figure}

\indent From the definition of $k_0$ in eq. \eqref{k0_Delta_r_perp_Definition} and the wave vector of the emitted radiation \eqref{WaveVectorCherenkov}, it is evident that the radiation will then be emitted almost paralell to the beam velocity. As a result, the Cherenkov radiation emitted through the mantle of the cylinder will no longer dominate the total emission of Cherenkov radiation.\\
\indent Therefore we now consider Cherenkov radiation emitted through the top of the same cylinder with radius $R_c$ and length $L_c$. Suppose there is some circular detector with inner and outer radii $R_1 < R_2 = R_c$, as is shown in fig. \ref{Figure1}. If we require, similarly to section \ref{SubsectionMantleSpectrum}, that the inner radius $R_1$ is much larger than the beam radius $R_b$ and also chosen in the far-field regime, i.e. $\vert k_0 R_1\vert \gg 1$, we can again use the far-field expressions \eqref{E_omega_approx_result} for the fields and integrate over the ring-shaped cutout of the cylinder top surface, which we denote by $T_c$.\\
\indent Defining

\begin{equation}
\xi_2(\omega) = \frac{c^2}{v^2} \frac{\text{Re}(\varepsilon)}{\vert\varepsilon\vert^2} \left\vert \sqrt{\frac{\varepsilon v^2}{c^2} - 1}\right\vert\,,
\label{dE_domega_top_help}
\end{equation}

\noindent it follows (see appendix)

\begin{equation}
\frac{d\mathcal{E}}{d\omega dr_{\perp}}\Biggl\vert_{T_c} \approx \frac{e^2}{c^2} \, \omega \, \xi_2(\omega) e^{-2\text{Im}(k_0)r_{\perp}} \vert S\vert^2\,.
\label{dE_domega_top_result1.5}
\end{equation}

\noindent If absorption on the scale of $r_{\perp} \in [R_1,R_2]$ is negligible, i.e. $\text{Im}(k_0) R_2 \ll 1$, we can write

\begin{equation}
\begin{aligned}
\frac{d^2\mathcal{E}}{d\omega dr_{\perp}}\Biggl\vert_{T_c} &\approx \frac{e^2}{c^2} \, \omega \, \xi_2(\omega)\, \vert S\vert^2\,.
\end{aligned}
\label{dE_domega_top_result2}
\end{equation}

\noindent Completely neglecting absorption leads to the simple result

\begin{equation}
\frac{d^2\mathcal{E}}{d\omega dr_{\perp}}\Biggl\vert_{T_c} = \frac{e^2}{c^2} \, \omega \, \frac{c^2}{\varepsilon v^2} \sqrt{\frac{\varepsilon v^2}{c^2} - 1}\, \vert S\vert^2\,.
\label{dE_domega_top_result3}
\end{equation}

\end{subsection}

\begin{subsection}{Coherence Factor and Angular Relations}\label{SubsectionCoherenceFactor_AngularRelations}
Let us now discuss two perspectives on the coherence factor, first a more physically motivated one and secondly a more general statistical picture:\\
\indent Using the definition of $S$ in eq. \eqref{S_Definition} we can define coherence lengths in the perpendicular and parallel direction to the beam-electron velocity \cite{CohImpactIonization}

\begin{equation}
\begin{aligned}
\lambda_{\perp} &:= \frac{1}{\text{Re}(k_0)} \quad, \quad\lambda_{\parallel} &:= \frac{v}{\omega}\,.
\end{aligned}
\label{CoherenceLenghts_Definition}
\end{equation}

\noindent Let $\Delta l_{\perp},\Delta l_{\parallel}$ be the mean interparticle distances in the electron beam in the perpendicular and parallel direction relative to its velocity. We can then interpret the coherence lengths as "resolutions" of the process with respect to the electrons in the beam. Thus, if we require the beam to be dense in comparison to the respective coherence lengths, i.e.

\begin{equation}
\begin{aligned}
\Delta l_{\perp} &\ll \lambda_{\perp} \quad ,\quad \Delta l_{\parallel} &\ll\,, \lambda_{\parallel}\,,
\end{aligned}
\label{PhaseFactor_IntegralCondition}
\end{equation}

\noindent we can perform a transition from a sum to an integral in $S$, i.e.

\begin{equation}
S \rightarrow \int_{\mathbb{R}^3} d^3{\bf R} \, \varrho({\bf R}) e^{-i{\bf k}_{\perp,0} \cdot {\bf R}_{\perp}} e^{-i\frac{\omega Z}{v}}\,,
\label{PhaseFactor_TransitionToIntegral}
\end{equation}

\noindent where $\varrho({\bf R})$ is the particle density and ${\bf R} = ({\bf R}_{\perp}, Z)$. Note the change in the sign of the exponential function corresponding to the previous factor $e^{i\omega t_\ell}$ because of the relation $vt_\ell = -(z_\ell - vt)$.\\
\indent Physically, for a certain frequency $\omega$, if both mean interparticle distances in the beam are much smaller than their respective coherence lengths, many beam-electron contributions to Cherenkov radiation with frequency $\omega$ will constructively interfere.\\
\indent More generally, for $N_e \gg 1$, we can view $S$ in a statistical picture and calculate the expectation value of $\vert S\vert^2$ as

\begin{align*}
\left\langle\vert S\vert^2\right\rangle &= \left\langle \sum_{n,m = 1}^{N_e} e^{-i{\bf k}_{\perp,0} \cdot ({\bf r}_{\perp,n} - {\bf r}_{\perp,m})} e^{i\omega(t_n - t_m)} \right\rangle\\
&= N_e + \left\langle \sum_{n\neq m} e^{-i{\bf k}_{\perp,0} \cdot ({\bf r}_{\perp,n} - {\bf r}_{\perp,m})} e^{i\omega(t_n - t_m)} \right\rangle\,.
\end{align*}

\noindent Assuming that all particle positions are independent variables distributed according to the same probability distribution and using the same argument for the sign change in the second exponential factor as in eq. \eqref{PhaseFactor_TransitionToIntegral} we thus arrive at

\begin{equation}
\begin{aligned}
\left\langle\vert S\vert^2\right\rangle &= N_e + N_e(N_e-1)\, \Pi\\
&= N_e\left[1 + (N_e - 1)\,\Pi\right]\,,
\end{aligned}
\label{PhaseFactorGeneralFormula}
\end{equation}

\noindent with the form factor

\begin{equation}
\Pi = \left\vert\left\langle e^{-i{\bf k}_{\perp,0} \cdot {\bf R}_{\perp}} e^{-i\frac{\omega}{v}Z} \right\rangle\right\vert^2
\label{FormFactor}
\end{equation}

\indent Explicitly, the form factor $\Pi$ is given by the absolute value squared of the integral in eq. \eqref{PhaseFactor_TransitionToIntegral}, divided by a factor $N_e^2$, since $\Pi$ is defined statistically through a probability distribution instead of a particle distribution. This in particular confirms the condition \eqref{PhaseFactor_IntegralCondition} for coherent enhancement in terms of coherence lengths, since the assumption $\Pi \gg 1/N_e$ in eq. \eqref{PhaseFactorGeneralFormula} leads back to eq. \eqref{PhaseFactor_TransitionToIntegral}.\\
\indent We consider three examples for the beam shape (i) a "doubly" Gaussian beam (ii) a cylindric constant-density beam (iii) a "doubly" super-Gaussian beam. Using the same notations $R_b, L_b$ for the effective beam radius and length as before, we define the beam particle densities

\begin{equation}
\begin{aligned}
\varrho_g(r_{\perp}, \eta) &= \frac{2N}{\pi^{3/2}L_b R_b^2} \, e^{-\left(\frac{\eta}{L_b/2}\right)^2} \, e^{-\left(\frac{r_{\perp}}{R_b}\right)^2}\,,\\
\varrho_c(r_{\perp}, \eta) &= \begin{cases} \frac{N_e}{\pi L_b R_b^2} \; ,\; \text{for} \; r_{\perp} \leq R_b\; , \; \vert\eta \vert\leq \frac{L_b}{2};\\ 0\; , \; \text{else.} \end{cases}\\
\varrho_{g_\nu}(r_{\perp}, \eta) &= \frac{2\nu^2 N_e}{\pi L_b R_b^2 \, \Gamma\left(\frac{1}{2\nu}\right) \Gamma\left(\frac{1}{\nu}\right)} \, e^{-\left(\frac{\eta}{L_b/2}\right)^{2\nu}} \, e^{-\left(\frac{r_{\perp}}{R_b}\right)^{2\nu}}\,,
\end{aligned}
\label{rho_definitions}
\end{equation}

\noindent where we have denoted $\eta := z - vt$. The corresponding phase factors are calculated from eq. \eqref{PhaseFactor_TransitionToIntegral} (see appendix), resulting in \cite{AbrStegun}:

\begin{equation}
\begin{aligned}
\Pi_g &= \exp\left(-\frac{1}{2}\frac{\omega^2}{v^2}\left(\left[\frac{\text{Re}(\varepsilon) v^2}{c^2} - 1\right] R_b^2 + \left(\frac{L_b}{2}\right)^2\right)\right)\,,\\
\Pi_c &= \left[\frac{\sin\left(\frac{\omega}{v}\frac{L_b}{2}\right)}{\frac{\omega}{v}\frac{L_b}{2}}\right]^2 \, \left\vert\frac{J_1\left(\frac{\omega R_b}{v} \sqrt{\frac{\varepsilon v^2}{c^2} - 1}\right)}{\frac{1}{2} \frac{\omega R_b}{v} \sqrt{\frac{\varepsilon v^2}{c^2} - 1}}\right\vert^2\,,\\
\Pi_{g_\nu} &= \left[ \sum_{\ell_1 = 0}^{\infty} \frac{(-1)^{\ell_1} (\frac{\omega L_b}{2v})^{2\ell_1}}{(2\ell_1)!} \frac{\Gamma\left(\frac{2\ell_1 + 1}{2\nu}\right)}{\Gamma\left(\frac{1}{2\nu}\right)} \right]^2\\
&\times\,\left\vert \sum_{\ell_2 = 0}^{\infty} \frac{(-1)^{\ell_2} \left(\frac{\omega R_b}{2v} \sqrt{\frac{\varepsilon v^2}{c^2} - 1}\right)^{2\ell_2}}{(\ell_2!)^2} \frac{\Gamma\left(\frac{\ell_2 + 1}{\nu}\right)}{\Gamma\left(\frac{1}{\nu}\right)} \right\vert^2\,.
\end{aligned}
\label{PhaseFactors_Pi_Gaussian&Constant}
\end{equation}

\indent Now we stress a few points:\\
First, it should be noted that there is a sizeable difference between the beam shapes in the scaling for large $R_b$ or $L_b$ or similarly for large $\omega$ (with some additional care to still fulfill the Cherenkov condition \eqref{CherenkovCondition} with $\varepsilon(\omega)$). As seen from the first line in eq. \eqref{PhaseFactors_Pi_Gaussian&Constant}, for a Gaussian beam, $\Pi_g$ decreases exponentially, while for a constant-density beam, assuming small absorption,

\begin{align*}
\Pi_c \sim \frac{1}{\omega^5 L_b^2 R_b^3 \left[\frac{\varepsilon v^2}{c^2} - 1\right]^{3/2}}\,,
\end{align*}

\noindent for $\frac{\omega R_b}{v}, \, \frac{\omega L_b}{v} \gg 1$.\\
\indent Secondly, highly relativistic electron beams typically have much larger radius $R_b$ than length $L_b$ when they arrive at their targets \cite{PlasmaBeams5}. In order to improve transverse coherence, one should therefore choose a medium which has $\left(\text{Re}(\varepsilon) - 1\right) \ll 1$ to counteract this effect.\\
\indent Thirdly, the definition \eqref{FormFactor} of the form factor $\Pi$ is essentially the absolute value squared of the particle distribution's Fourier transform evaluated at the wave vector $\boldsymbol{k} = -\boldsymbol{\kappa}$. Consequently, faster changes in the beam density enable coherent enhancement for higher frequencies $\omega$. "Fast" in this context is determined by comparing the dimensions of the beam density fluctuations to the $\omega$-dependent coherence lengths $(\lambda_{\perp},\lambda_{\parallel})$. This we will confirm later in section \ref{SectionResults}.\\
\indent Finally we can interpret the previous results using the Cherenkov angle $\vartheta$. Using the definition \eqref{WaveVectorCherenkov} of the Cherenkov wave vector $\boldsymbol{\kappa}$, we see that the emission angle $\vartheta_{\kappa}$ is given by

\begin{align*}
\tan(\vartheta_{\kappa}) = \frac{\text{Re}(\kappa_{\perp})}{\kappa_z} = \text{Re}\left(\sqrt{\frac{\varepsilon v^2}{c^2} - 1}\right) = \tan(\vartheta)\,.
\end{align*}

\noindent For small absorption, i.e. $\frac{\text{Re}(\varepsilon)v^2}{c^2} - 1 \gg \text{Im}(\varepsilon)$, we can approximate $\text{Re}\left(\sqrt{\frac{\varepsilon v^2}{c^2} - 1}\right) \approx \sqrt{\frac{\text{Re}(\varepsilon)v^2}{c^2} - 1}$ and define the Cherenkov angle in the more commonly known form as

\begin{equation}
\cos(\vartheta) \approx \frac{c}{v \sqrt{\text{Re}(\varepsilon)}}\,.
\label{CherenkovAngle}
\end{equation}

\noindent Equivalently, we also have

\begin{align*}
\sqrt{\frac{\varepsilon v^2}{c^2} - 1} \approx \tan(\vartheta)\;,\;1 - \frac{c^2}{\varepsilon v^2} \approx \sin^2(\vartheta)\,.
\end{align*}

\indent With the reduced photon wavelength $\lambdabar = \frac{c}{n\omega}$, we can then write the coherence lengths as

\begin{align*}
\lambda_\parallel \approx \frac{\lambdabar}{\cos(\vartheta)}\;,\; \lambda_\perp \approx \frac{\lambdabar}{\sin(\vartheta)}\,.
\end{align*}

\noindent The spectra in eqs. \eqref{dE_domega_dz_mantle_result3} and \eqref{dE_domega_top_result3} can be written in terms of the Cherenkov angle as

\begin{equation}
\begin{aligned}
\frac{d^2\mathcal{E}}{d\omega dz}\Biggl\vert_{M_c} &\approx \frac{e^2 \omega}{c^2} \vert S\vert^2 \sin^2(\vartheta)\,,\\
\frac{d^2\mathcal{E}}{d\omega dr_{\perp}}\Biggl\vert_{T_c} &\approx \frac{e^2 \omega}{c^2} \vert S\vert^2 \sin(\vartheta)\cos(\vartheta)\,.
\end{aligned}
\label{SpectraAngleExpression}
\end{equation}

\indent Concerning the question under which conditions a medium of finite size can be considered "infinite", we see in the calculated fields, that the Hankel-functions reach their asymptotic forms for $r_{\perp} \vert k_0\vert \gg 1$, i.e.

\begin{equation}
r_{\perp} \gg R_f = \frac{v}{\omega \left\vert\sqrt{\frac{\varepsilon v^2}{c^2} - 1}\right\vert} \approx \frac{v}{\omega\tan(\vartheta)}\,.
\label{AsympRadius}
\end{equation}

\noindent Here we defined the "formation radius" $R_f$. Evidently if for some $\omega$ we have $\left[R_f \, \text{Im}(k_0)\right] \gg 1$, Cherenkov radiation for the chosen frequency will be suppressed by absorption in the medium.\\
\indent Equivalently, we can define a longitudinal distance $L_f$ through Huygens principle. For this we imagine a charged particle emitting spherical waves which propagate through the medium at velocity $\frac{c}{\text{Re}(\varepsilon)}$, forming a wave front with the characteristic Cherenkov angle. We can then determine the length a charged particle needs to traverse through the medium for the spherical wave emitted from its starting point to contribute to the Cherenkov wavefront at transverse distance $R_f$. This leads to the definition of a "formation length" as

\begin{equation}
L_f = \frac{v}{\omega}\left(1 + \frac{1}{\left\vert\frac{\varepsilon v^2}{c^2} - 1\right\vert}\right) \approx \frac{v}{\omega\sin^2(\vartheta)}\,.
\label{AsympLength}
\end{equation}

\indent Now let us assume that we measure the emission of Cherenkov radiation in a certain frequency range $[\omega_1,\omega_2]$ which is far from absorption resonances in the medium, such that $\text{Im}(k_0) \ll \text{Re}(k_0)$. In that frequency range, we can determine a $\omega_0$ for which $\text{Re}(\varepsilon(\omega))$ becomes maximal, which gives the minimum Cherenkov angle

\begin{align*}
\vartheta_{\text{min}} = \text{arccos}\left(\frac{c}{\text{Re}(\varepsilon(\omega_0))v}\right)\,.
\end{align*}

\noindent A medium can then be considered "infinite" for this frequency range, if it is much larger than a cylinder of length and radius

\begin{align*}
L_{f,\text{max}} \approx \frac{v}{\omega_0 \sin^2(\vartheta_{\text{min}})}\;,\; R_{f,\text{max}} \approx \frac{v}{\omega_0\tan(\vartheta_{\text{min}})}\,.
\end{align*}

\indent Moreover, if we calculate the spectra at a certain radius $r_{\perp}$, this additionally assumes that the medium extends around the point on the beam propagation axis at longitudinal distance $\frac{r_{\perp}}{\tan(\vartheta)}$ from the detection point by many formation lengths $L_f$.\\
\indent Lastly, we consider the effects of energy spread in a highly relativistic electron beam on the coherent Cherenkov spectrum. Taking $\gamma = 2000$, a typical relative energy spread $\delta\gamma = \frac{\Delta\gamma}{\gamma}$ of 10\% \cite{PlasmaPhotocathode1} induces a relative spread $\delta\beta = \frac{\Delta\beta}{\beta} \sim 10^{-7}$ of the reduced velocity $\beta = \frac{v}{c}$.\\
\indent The single-electron spectrum of Cherenkov radiation depends on the beam energy only through the reduced velocity $\beta$. If the condition $(\text{Re}(\varepsilon)\beta^2 - 1) \gg 2\text{Re}(\varepsilon)\beta^2\,\delta\beta$ is fulfilled and absorption in the medium is small, we have $\sqrt{\varepsilon\beta^2(1 \pm 2\delta\beta) - 1} \approx \sqrt{\varepsilon\beta^2 - 1} \pm \frac{\varepsilon\beta^2\,\delta\beta}{\sqrt{\varepsilon\beta - 1}}$. For the typical value $(\text{Re}(\varepsilon)\beta^2 - 1) \approx 7.5\cdot 10^{-5}$ in section \ref{SectionResults}, we thus have $\frac{\varepsilon\beta^2\,\delta\beta}{\sqrt{\varepsilon\beta - 1}} \approx 10^{-5} \ll 10^{-2} \approx \sqrt{\text{Re}(\varepsilon)\beta^2 - 1}$ such that the incoherent spectrum is essentially constant over the whole velocity spread.\\
\indent Furthermore, for the coherence lengths we have $\delta\lambda_\parallel \approx \delta\beta$ and $\delta\lambda_\perp \approx \frac{\delta\beta}{\text{Re}(\varepsilon) \beta^2 - 1}$, which -- for the same parameters as taken above -- comes out to $\delta\lambda_\parallel \approx 10^{-7}$ and $\delta\lambda_\perp \approx 10^{-3}$. Then the absolute width of the ratios of beam dimensions $L_b$ and $R_b$ and the coherence lengths which are indicative of the beam coherence are given by $\frac{L_b}{\lambda_\parallel} \delta\lambda_\parallel$ and $\frac{R_b}{\lambda_\perp}\delta\lambda_\perp$, which both are much smaller than 1 for the beam parameters used in section \ref{SectionResults}. Only if $L_b \gtrsim 10^7 \lambda_\parallel$ or $R_b \gtrsim 10^3 \lambda_\perp$ could these deviances provide a signficant phase difference for the form factor, but then the beam would act completely incoherently.
\end{subsection}

\section{Results}\label{SectionResults}
\begin{subsection}{Dielectric Permittivity Model}
As shown in section \ref{SubsectionCoherenceFactor_AngularRelations}, it is advantageous for coherent radiation from a highly relativistic beam to choose a medium with $\left(\text{Re}(\varepsilon) - 1\right) \ll 1$. One such candidate for frequencies up to the low-ultraviolet regime is a helium gas target.\\
\indent To model the dielectric permittivity of helium, we use the Lorentz oscillator model \cite{Fließbach} which we formulate as

\begin{equation}
\varepsilon(\omega) = 1 + C\frac{4\pi n_0 e^2}{m_e} \sum_j \, \frac{f_j}{\omega_j^2 - \omega^2 - i\omega\Gamma_j/\hbar}\,,
\label{LorentzOscillator}
\end{equation}

\noindent with normalization factor $C$, atomic density $n_0$, electron mass $m_e$, atomic transition strengths $f_j$ corresponding to transition energies $\hbar\omega_j$ and decay widths $\Gamma_j$.

\begin{table}[H]
\centering
\begin{tabular}{r|c|c|c}
& $\hbar\omega_j \,[\text{eV}]$ & $\Gamma_j/\hbar \, [1/\text{s}]$ & $f_j$\\
\hline
$\text{X}=2$ & $\;21.21804\;$ & $\;1.7989 \cdot 10^9\;$ & $\;2.7625 \cdot 10^{-1}\;$\\
\hline
$\text{X}=3$ & $\;23.08705\;$ & $\;5.6634 \cdot 10^8\;$ & $\;7.3460 \cdot 10^{-2}\;$\\
\hline
$\text{X}=4$ & $\;23.7433\;$ & $\;2.4356 \cdot 10^8\;$ & $\;2.9873 \cdot 10^{-2}\;$\\
\hline
$\text{X}=5$ & $\;24.0468\;$ & $\;1.2582 \cdot 10^8\;$ & $\;1.5045 \cdot 10^{-2}\;$\\
\hline
$\text{X}=6$ & $\;24.212\;$ & $\;7.3174 \cdot 10^7\;$ & $\;8.6306 \cdot 10^{-3}\;$\\
\hline
$\text{X}=7$ & $\;24.312\;$ & $\;4.6224 \cdot 10^7\;$ & $\;5.4073 \cdot 10^{-3}\;$
\end{tabular}
\caption{Transition energies $\hbar\omega_j$, Einstein coefficients $\Gamma_j/\hbar$ and transition strengths $f_j$ taken from NIST Atomic Spectra Database for dipole transitions $1\text{s}^2 \rightarrow 1\text{s}\text{Xp}$ in helium \cite{NIST}.}
\label{Table_NIST}
\end{table}

\indent To obtain an accurate model for frequencies $\omega$ well below the helium ionization threshold ($\sim$24.6 eV), we take the 6 strongest bound-bound transitions starting from the helium ground state, i.e. the dipole transitions $1\text{s}^2 \rightarrow 1\text{s}\text{Xp}$ with $\text{X} = 2, \,3, \,4, \, 5, \, 6, \,7$. The data was obtained from the NIST Atomic Spectra Database \cite{NIST} and is shown in table \ref{Table_NIST}.\\
\indent In this model, transitions to highly excited states and the continuum have not been taken into account. For the frequencies well below the absorption resonances, this would lead to a slight underestimation of $\varepsilon$ compared to experimental values for a respective choice of $n_0$.\\
\indent For Helium at a temperature $273.15\,\text{K}$ and pressure $1\,\text{atm}$, i.e. with $n_0 \approx 2.687 \cdot 10^{19}\,\text{cm}^{-3}$, the experimentally measured value for the real part of the dielectric permittivity is $(\text{Re}(\varepsilon) - 1) = 6.90 \cdot 10^{-5}$ \cite{RefractiveIndex}. For the same $n_0$, eq. \eqref{LorentzOscillator} with $C=1$ gives $(\text{Re}(\varepsilon) - 1) \approx 6.33 \cdot 10^{-5}$. By comparison we can thus choose $C = \frac{6.90}{6.33} \approx 1.09$ to "normalize" the permittivity model. Since $(\text{Re}(\varepsilon) - 1)$ far below the first absorption resonances is linear in $n_0$ in very good approximation \cite{RefractiveIndex2}, the low frequency limit should then be accurate for other values of $n_0$ as well.

\begin{figure}[H]
\begin{subfigure}{\linewidth}
\centering
\includegraphics[width=8.8cm]{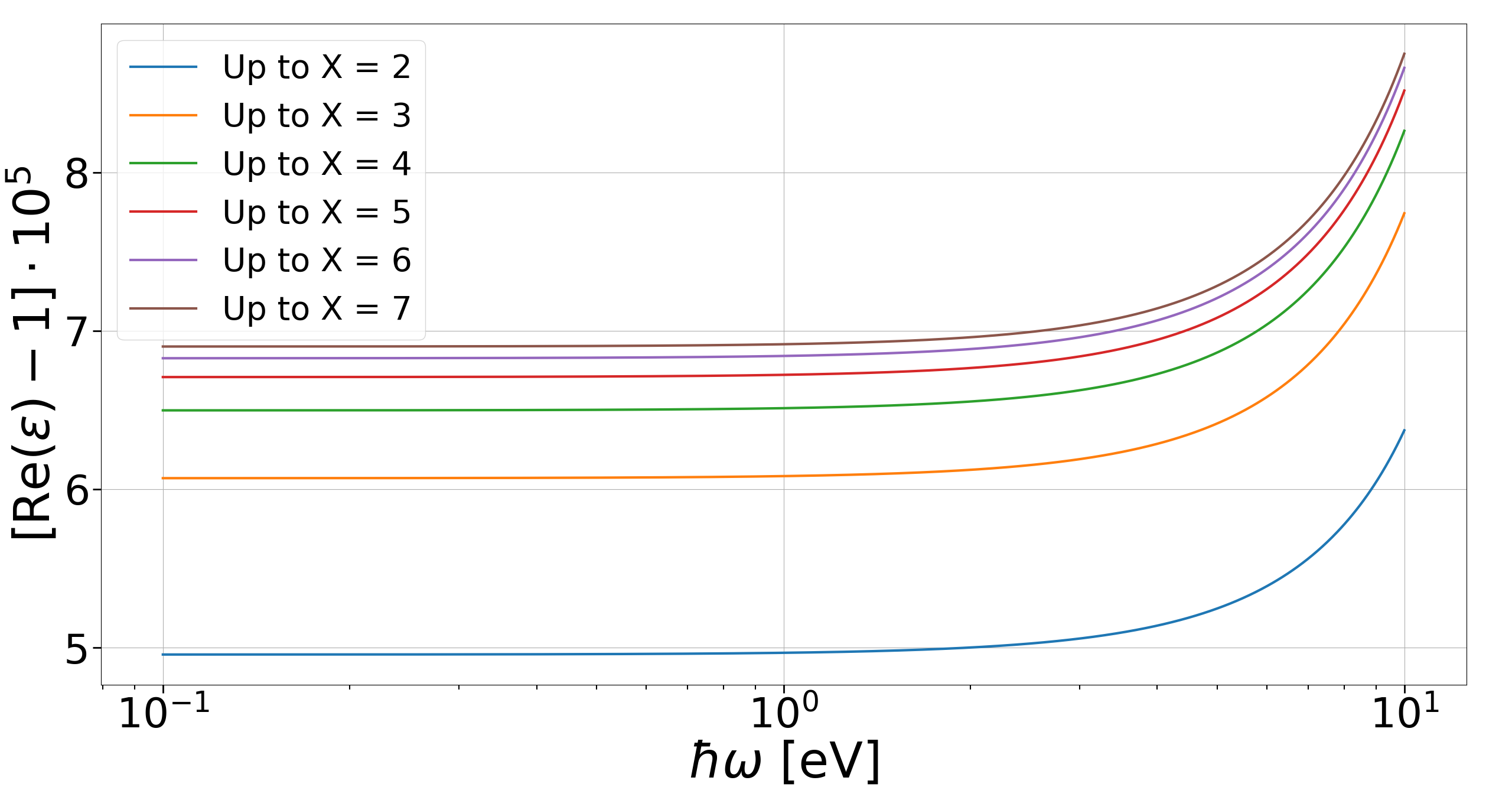}
\caption{}
\label{Figure2a}
\end{subfigure}
\begin{subfigure}{\linewidth}
\centering
\includegraphics[width=8.8cm]{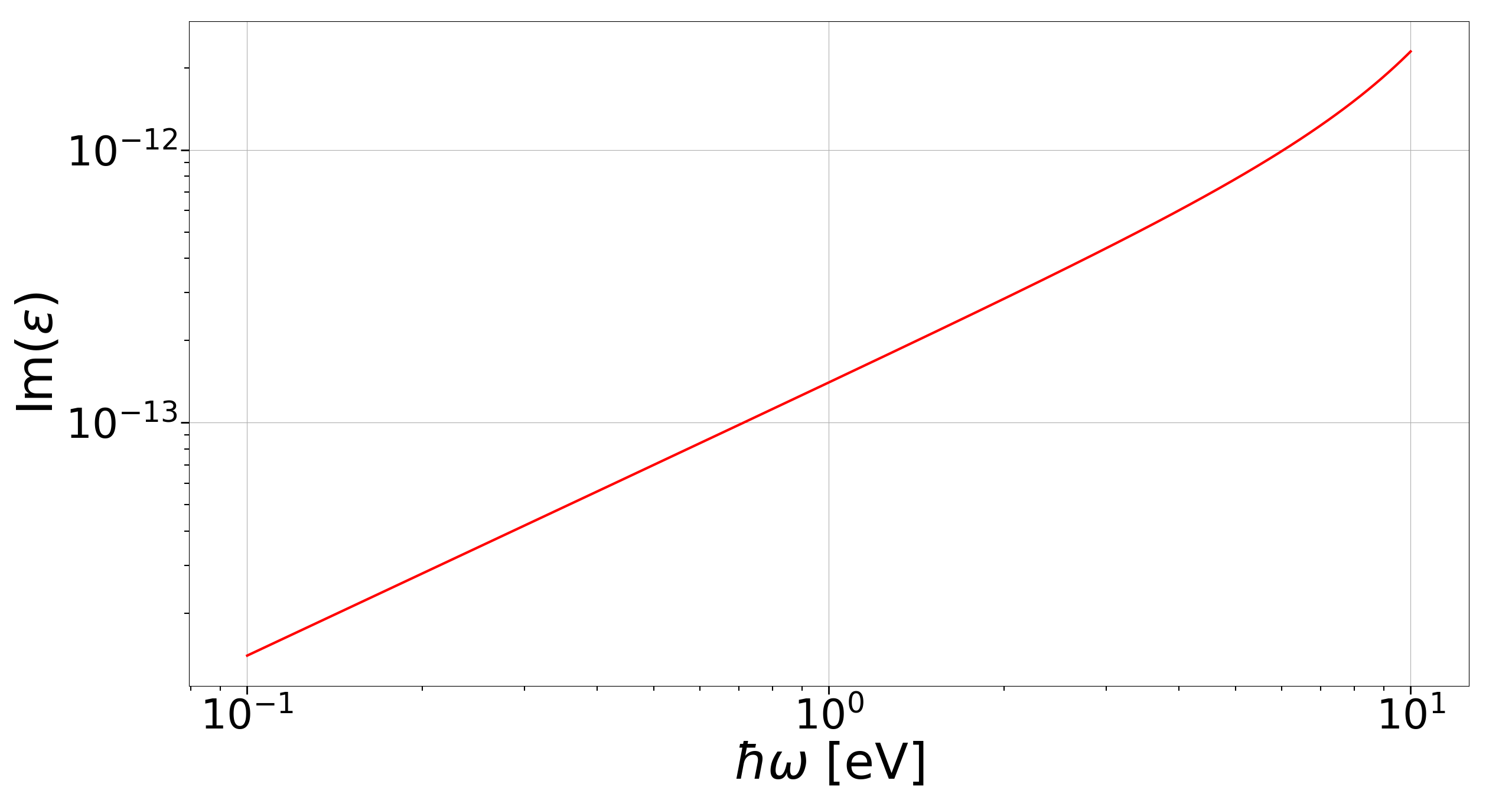}
\caption{}
\label{Figure2b}
\end{subfigure}
\caption{Dieletric permittivity $\varepsilon$ of helium as a function of photon energy as given in eq. \eqref{LorentzOscillator} with $n_0 = 2.687 \cdot 10^{19} \, \text{cm}^{-3},\,C = 1.09$ and the data from table \ref{Table_NIST}. (a) $\left(\text{Re}(\varepsilon) - 1\right)$ multiplied by $10^5$, with varying number of included transitions, (b) $\text{Im}(\varepsilon)$ with all 6 transitions included.}
\label{Figure2}
\end{figure}

\end{subsection}

\begin{subsection}{Coherent Cherenkov Spectra From a Gaussian Beam}
Using this helium permittivity model, which is plotted in fig. \ref{Figure2}, we now calculate the spectra, for which we compare the fully incoherent case, i.e. $\vert S\vert^2 = N_e$ to the coherently enhanced case, where $\vert S\vert^2$ is calculated using eq. \eqref{PhaseFactorGeneralFormula}.\\
\indent Figs. \ref{Figure3a} and \ref{Figure3b} show the "mantle" and "top" spectra, respectively taken at $R_c = r_{\perp} = 1 \, \text{mm}$, for a compact Gaussian electron beam of about $1\,\text{GeV}$ energy penetrating a helium gas with density $n_0 = 2.687 \cdot 10^{19} \, \text{cm}^{-3}$, beam-electron energy of about $1 \,\text{GeV}$. A quick estimate shows that this is in the far-field regime, since $\left\vert k_0 r_{\perp} \right\vert \gtrsim 4$, where $\omega = 10^{-1} \, \text{eV}$ was taken for this lower bound. As $\text{Im}(\varepsilon)$ is at least 7 orders of magnitude smaller than $\left(\text{Re}(\varepsilon) - 1\right)$ for $\hbar\omega \lesssim 10\,\text{eV}$ (see fig. \ref{Figure2}), we can also assess from the same estimate that absorption is negligible for all following numerical results.

\begin{figure}[H]
\centering
\begin{subfigure}{\linewidth}
\includegraphics[width=8.8cm]{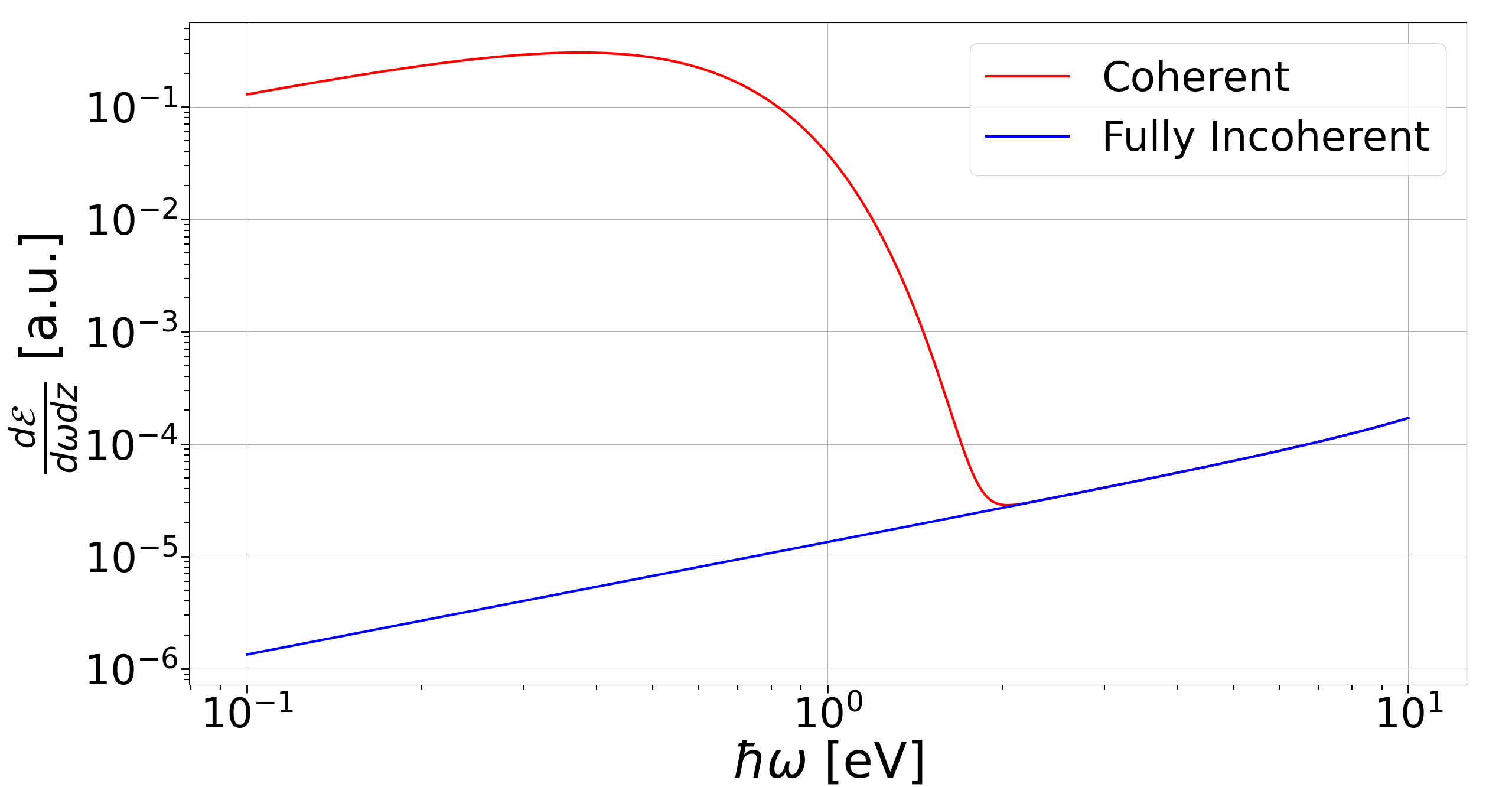}
\caption{}
\label{Figure3a}
\end{subfigure}
\begin{subfigure}{\linewidth}
\includegraphics[width=8.8cm]{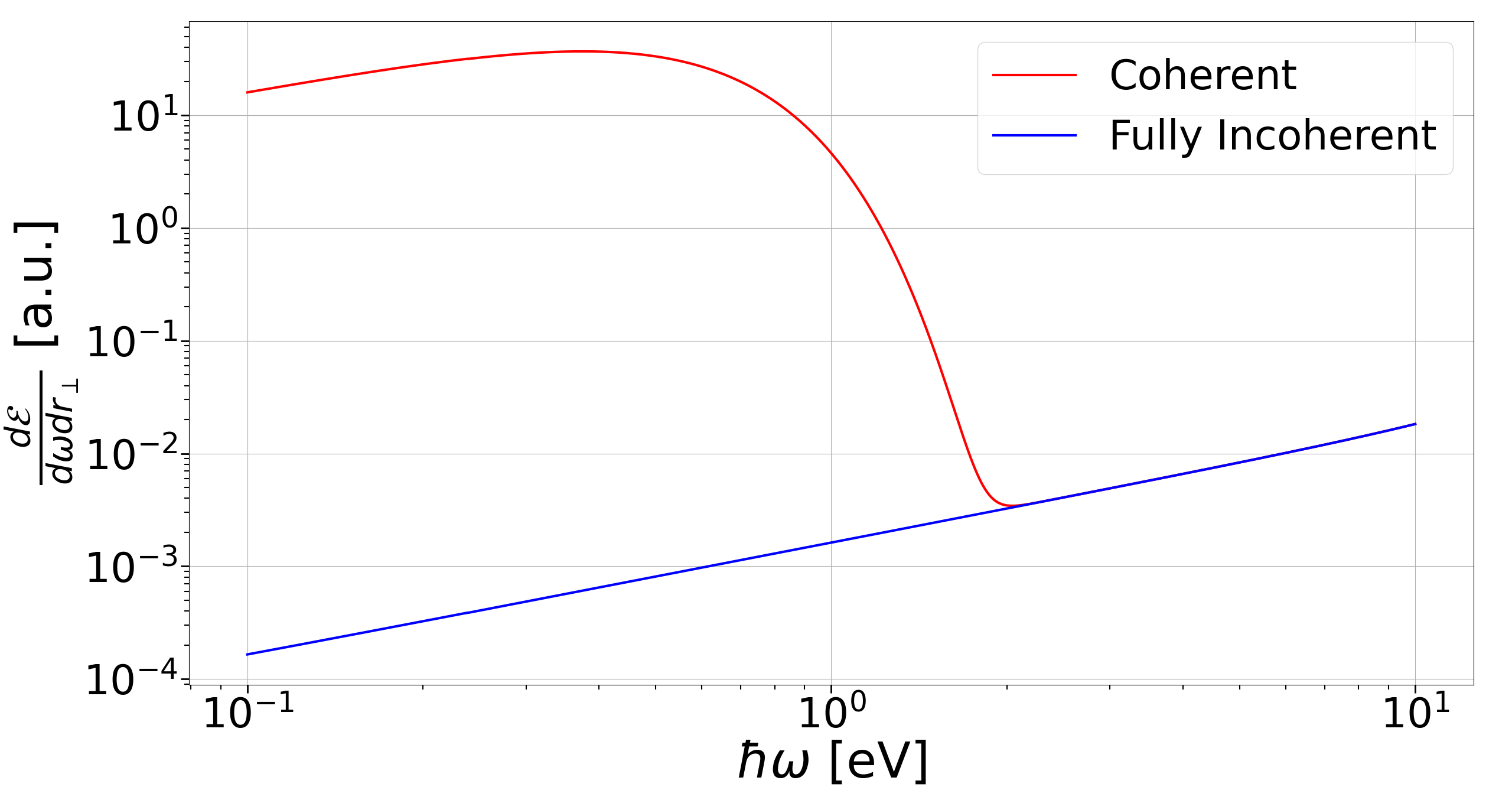}
\caption{}
\label{Figure3b}
\end{subfigure}
\caption{Cherenkov energy spectrum by a Gaussian beam as defined in eq. \eqref{rho_definitions}, as a function of photon energy. Parameters are $n_0 = 2.687 \cdot 10^{19} \, \text{cm}^{-3}, \,R_c = r_{\perp} = 1 \,\text{mm}, \, N_e = 10^5, \,\gamma = 2000, \, R_b = 20 \, \mu\text{m}, \, L_b = 1 \, \mu\text{m}$. "Fully Incoherent" case sets $\vert S\vert^2 = N_e$, whereas "Coherent" case utilizes full expression \eqref{PhaseFactorGeneralFormula} for $\vert S\vert^2$. (a) Spectrum through mantle of cylindrical volume ("mantle spectrum"), (b) Spectrum through top of cylindrical volume ("top spectrum").}
\label{Figure3}
\end{figure}

\indent The linear dependence of the fully incoherent spectra on $\omega$ is well visible, as well as coherent enhancement up to photon energies of about $2 \, \text{eV}$.\\
\indent Evidently, the top spectrum is 2 orders of magnitude larger than the mantle spectrum, which is due to the small Cherenkov angle, see eq. \eqref{SpectraAngleExpression}. We can write

\begin{align*}
\frac{\text{Re}(\varepsilon) v^2}{c^2} - 1 = \text{Re}(\varepsilon) - 1 - \frac{\text{Re}(\varepsilon)}{\gamma^2} \,,
\end{align*}

\noindent from which we obtain

\begin{align*}
\vartheta \approx \tan(\vartheta) \approx \sqrt{\text{Re}(\varepsilon) - 1 - \frac{\text{Re}(\varepsilon)}{\gamma^2}}\,.
\end{align*}

\noindent Inserting $\text{Re}(\varepsilon) - 1 \approx 7.5 \cdot 10^{-5}$ as a typical value and $\gamma = 2000$, this yields $\vartheta \approx 8.65 \cdot 10^{-3}$. Taking the inverse, this implies a factor of about $116$ between the mantle and top spectra, which is in good agreement with figures \ref{Figure3a} and \ref{Figure3b}.\\
\indent The mantle spectrum could in principle be measured by putting a cylindric mantle detector around the beam trajectory. For the $\omega$-interval chosen in fig. \ref{Figure7}, such a cylindric detector, for the parameters chosen here, would have to be about $1/\vartheta \approx 116$ times as long as the difference of outer and inner radius $(R_2 - R_1)$ of a ring detector as shown in fig. \ref{Figure1} to measure the same amount of photons as the ring detector. One should keep in mind that this estimate assumes, according to section \ref{SectionTheory}, a medium long enough such that radiation emitted at the Cherenkov angle would arrive over the full length of that cylindric mantle detector. Since the mantle spectrum in the far-field zone for low absorption is proportional to the top spectrum, all considerations concerning the shape of the top spectrum are valid for the mantle spectrum as well.\\
\indent In the following, we will thus neglect the mantle spectrum since it is heavily suppressed through the small Cherenkov angle.\\
\indent To have coherent enhancement for a Gaussian beam, we demand $\Pi_g \gtrsim \frac{1}{N_e}$ and use eq. \eqref{PhaseFactors_Pi_Gaussian&Constant} to formulate the implicit condition

\begin{equation}
\omega \lesssim \sqrt{\frac{2v^2\ln(N_e)}{\left[\frac{\text{Re}(\varepsilon(\omega)) v^2}{c^2} - 1\right]R_b^2 + \left(\frac{L_b}{2}\right)^2}}\,,
\label{CoherenceCondition_Gaussian}
\end{equation}

\noindent which -- for the parameters chosen in fig. \ref{Figure3} and taking the maximal $\frac{\text{Re}(\varepsilon(\omega)) v^2}{c^2} - 1 \approx 8.5 \cdot 10^{-5}$ in our chosen $\omega$-interval -- comes out to $\hbar\omega \lesssim 1.8 \, \text{eV}$, which is reflected within the plots.

\end{subsection}

\begin{subsection}{Coherent Cherenkov Spectra From a Cylindric Constant-Density Beam}
\indent Let us now compare with the case of a cylindric electron beam of constant density. To this end, the cylindrical constant-density beam is assumed to have the same parameter values $N_e = 10^5,\,L_b = 1\,\mu\text{m},\,R_b = 20\,\mu\text{m}$ and $\gamma = 2000$ as were taken for the Gaussian beam. Comparing with the case of a Gaussian beam where coherent enhancement extends up to about $\hbar \omega \approx 2 \, \text{eV}$, figure \ref{Figure4} shows coherent enhancement for much higher frequencies $\omega$, up to about $\hbar\omega \approx 10 \, \text{eV}$. Physically, this is understandable since the sudden changes in the beam density at the edges of the constant-density cylinder lead to higher-frequency components in the coherence factor.

\begin{figure}[H]
\includegraphics[width=8.8cm]{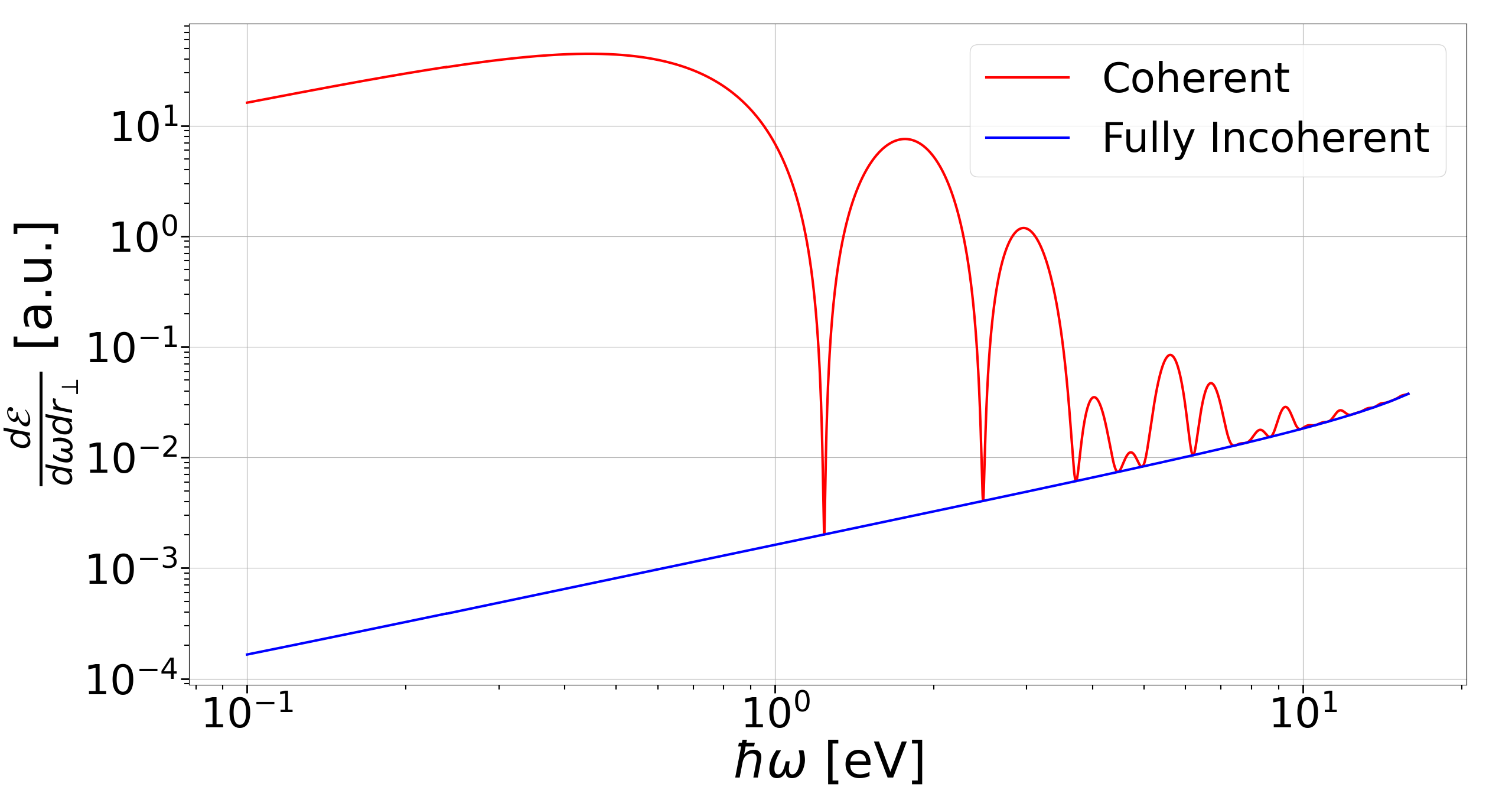}
\caption{Cherenkov energy spectrum ("top spectrum") by a cylindric constant-density beam as defined in eq. \eqref{rho_definitions}, as a function of photon energy. Parameters are $n_0 = 2.687 \cdot 10^{19} \, \text{cm}^{-3}, \,r_{\perp} = 1 \,\text{mm}, \, N_e = 10^5, \,\gamma = 2000, \, R_b = 20 \, \mu\text{m}, \, L_b = 1 \, \mu\text{m}$. "Fully Incoherent" case sets $\vert S\vert^2 = N_e$, whereas "Coherent" case utilizes full expression \eqref{PhaseFactorGeneralFormula}.}
\label{Figure4}
\end{figure}

\indent Additionally, the oscillatory nature of $\Pi_c$ from eq. \eqref{PhaseFactors_Pi_Gaussian&Constant} is clearly visible. The oscillation periods for longitudinal and transverse coherence can be written as

\begin{equation}
\begin{aligned}
\Delta\omega_{\parallel} = \frac{2\pi v}{L_b}\,, \quad \Delta\omega_{\perp} = \frac{\pi v}{R_b\,\text{Re}\left(\sqrt{\frac{\varepsilon v^2}{c^2} - 1}\right)}\,.
\end{aligned}
\label{OscillationPeriods}
\end{equation}

\noindent For the parameters chosen in fig. \ref{Figure4}, these oscillation periods come out to $\Delta\omega_{\parallel}\approx 1.24 \, \text{eV}$ and $\Delta\omega_{\perp}\approx 3.66 \, \text{eV}$. This is in good agreement with fig. \ref{Figure4}, as there are oscillations in the spectrum with oscillation period of $\sim 1\, \text{eV}$ until $\hbar\omega \sim 3.5 \, \text{eV}$, where the oscillations become irregular since both transverse and longitudinal coherence are reduced, leading to mixed oscillations of the $\sin^2$ and the Bessel function term in $\Pi_c$. Physically, these oscillations arise because a "plateau" in the beam density effectively acts as a resonance length which causes destructive interference when a multiple of the oscillation period of emitted radiation is equal to the time the beam needs to traverse the plateau length.\\
\indent Requiring $\Pi_c \gtrsim 1/N_e$ as the condition for coherent enhancement, we have to distinguish different cases:\\
First, let us assume that $\frac{L_b}{2}$ and $R_b\, \text{Re}\left(\sqrt{\frac{\varepsilon v^2}{c^2} - 1}\right)$ are within the same order of magnitude, which we denote as case I). Then both the $\text{sinc}^2$ and the Bessel function parts of $\Pi_c$ contribute to the decrease in coherence and we can insert asymptotic expansions for the $\text{sinc}^2$ and the Bessel function into eq. \eqref{PhaseFactors_Pi_Gaussian&Constant} to obtain \cite{AbrStegun}

\begin{align*}
\omega^5 \lesssim \frac{32N_e v^5}{\pi} \, \frac{\sin^2\bigl(\frac{\omega}{v}\frac{L_b}{2}\bigr) \cos^2\Bigl(R_b\, \text{Re}(k_0) - \frac{3\pi}{4}\Bigr)}{L_b^2 R_b^3 \left[\text{Re}\left(\frac{\varepsilon v^2}{c^2} - 1\right)\right]^{3/2}}.
\end{align*}

\noindent Approximating the product $\sin^2(..)\cos^2(..)$ as $\frac{1}{4}$ (see appendix), we obtain the condition

\begin{equation}
\text{I})\quad\omega \lesssim \sqrt[5]{ \frac{8N_e v^5}{\pi L_b^2 R_b^3} \left[\text{Re}\left(\frac{\varepsilon v^2}{c^2} - 1\right)\right]^{3/2}}\,.
\label{CoherenceCondition_Cylinder_Case1}
\end{equation}

\indent If either $\frac{L_b}{2} \gg R_b\, \text{Re}\left(\sqrt{\frac{\varepsilon v^2}{c^2} - 1}\right)$, case II), or $\frac{L_b}{2} \ll R_b\, \text{Re}\left(\sqrt{\frac{\varepsilon v^2}{c^2} - 1}\right)$, case III), only the $\text{sinc}^2$ term or the Bessel function term in $\Pi_c$ will limit coherence respectively. Approximating $\sin^2(...)$ or $\cos^2(...)$ as $\frac{1}{2}$ in these two cases, we obtain

\begin{equation}
\begin{aligned}
\text{II})&\quad \omega \lesssim \sqrt{2N_e}\frac{v}{L_b}\,,\\
\text{III})&\quad \omega \lesssim \sqrt[3]{\frac{4N_e}{\pi}} \frac{v}{R_b\, \text{Re}\left(\sqrt{\frac{\varepsilon v^2}{c^2} - 1}\right)}\,.
\end{aligned}
\label{CoherenceCondition_Cylinder_Case2_Case3}
\end{equation}

\indent For the parameters used in fig. \ref{Figure4}, we have $L_b/2 = 0.5 \, \mu\text{m}$ and, again taking the typical value $\text{Re}(\varepsilon) - 1 \approx 7.5 \cdot 10^{-5}$, $R_b\, \text{Re}\left(\sqrt{\frac{\varepsilon v^2}{c^2} - 1}\right) \approx 0.17 \, \mu\text{m}$, which is in between cases I) and II), where for case I) the limit would be $\omega\lesssim 6.8 \, \text{eV}$ and for case II) it would be $\omega \lesssim 88 \, \text{eV}$. The coherent enhancement up to about $10 \, \text{eV}$ is thus in agreement with the limits derived above.\\
\indent At the limit $88\,\text{eV}$ in case II), the Cherenkov condition can not be fulfilled since $\text{Re}(\varepsilon) < 1$ for high frequencies above the ionization threshold. Consequently the actual maximum coherently enhanced frequency in case II) would coincide with the first cutoff due to the Cherenkov condition which lies much lower near the first absorption resonance around $21.22\,\text{eV}$.\\
\indent Eq. \eqref{CoherenceCondition_Gaussian} implies that, for large $N_e$, the logarithmic scaling in $N_e$ for the Gaussian beam will make a sizeable difference in the maximum coherently enhanced $\omega$ compared to the constant-density cylinder beam. In a sense, these two scenarios represent extreme examples. The more realistic Gaussian distribution varies "smoothly" and thus does not contain many high-frequency components in its Fourier-transform, while the constant-density beam has sharp edges and thereby still supports coherent emission for much higher frequencies.
\end{subsection}
\subsection{Coherent Cherenkov Spectra From a Super-Gaussian Beam}
\begin{figure}[H]
\includegraphics[width=8.8cm]{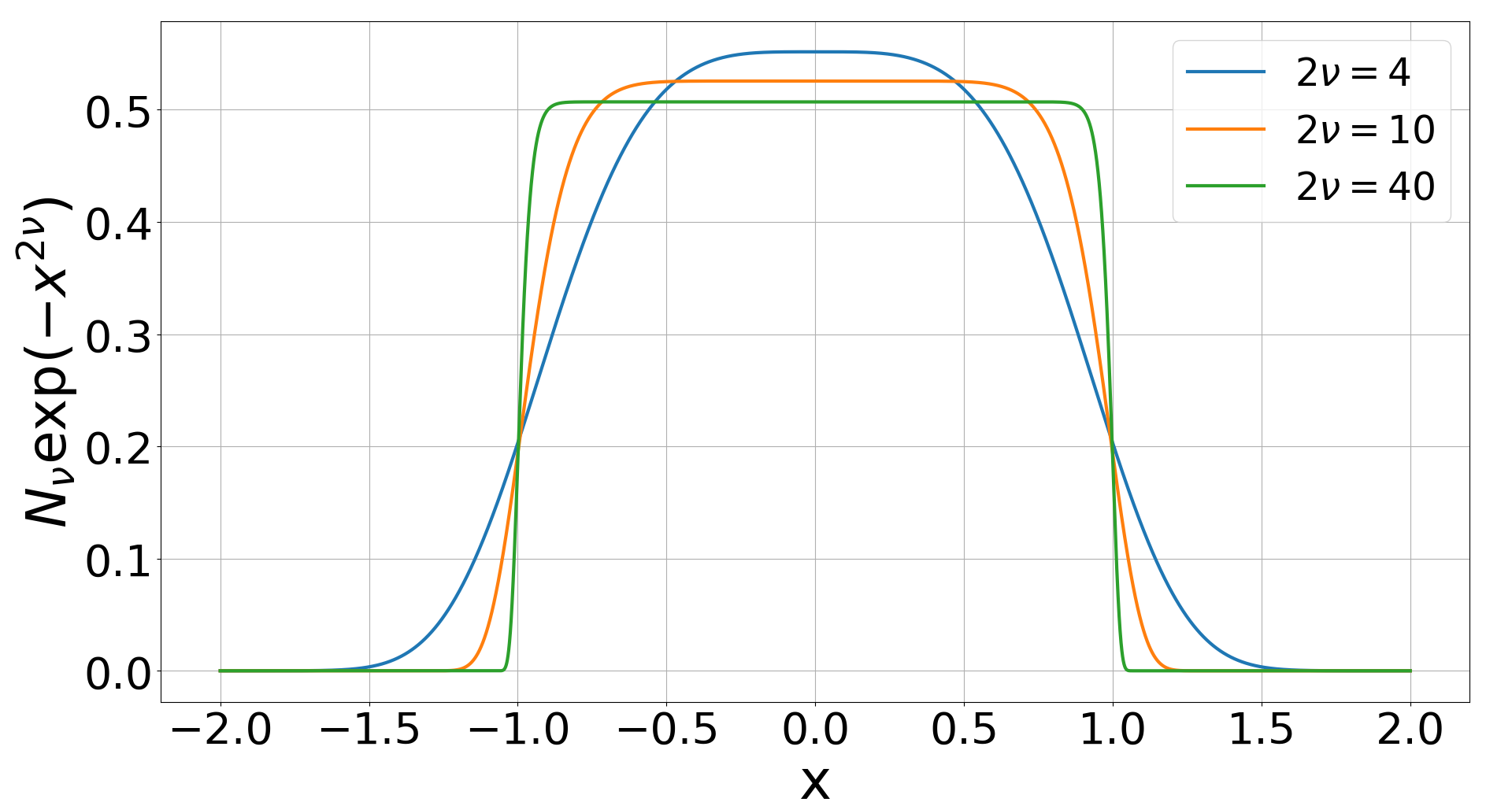}
\caption{One-dimensional super-Gaussian distributions $\mathcal{N}_\nu \exp\left(-x^{2\nu}\right)$ normalized to $1$, for $\nu=2,5,20$.}
\label{Figure5}
\end{figure}

\begin{figure}[H]
\includegraphics[width=8.8cm]{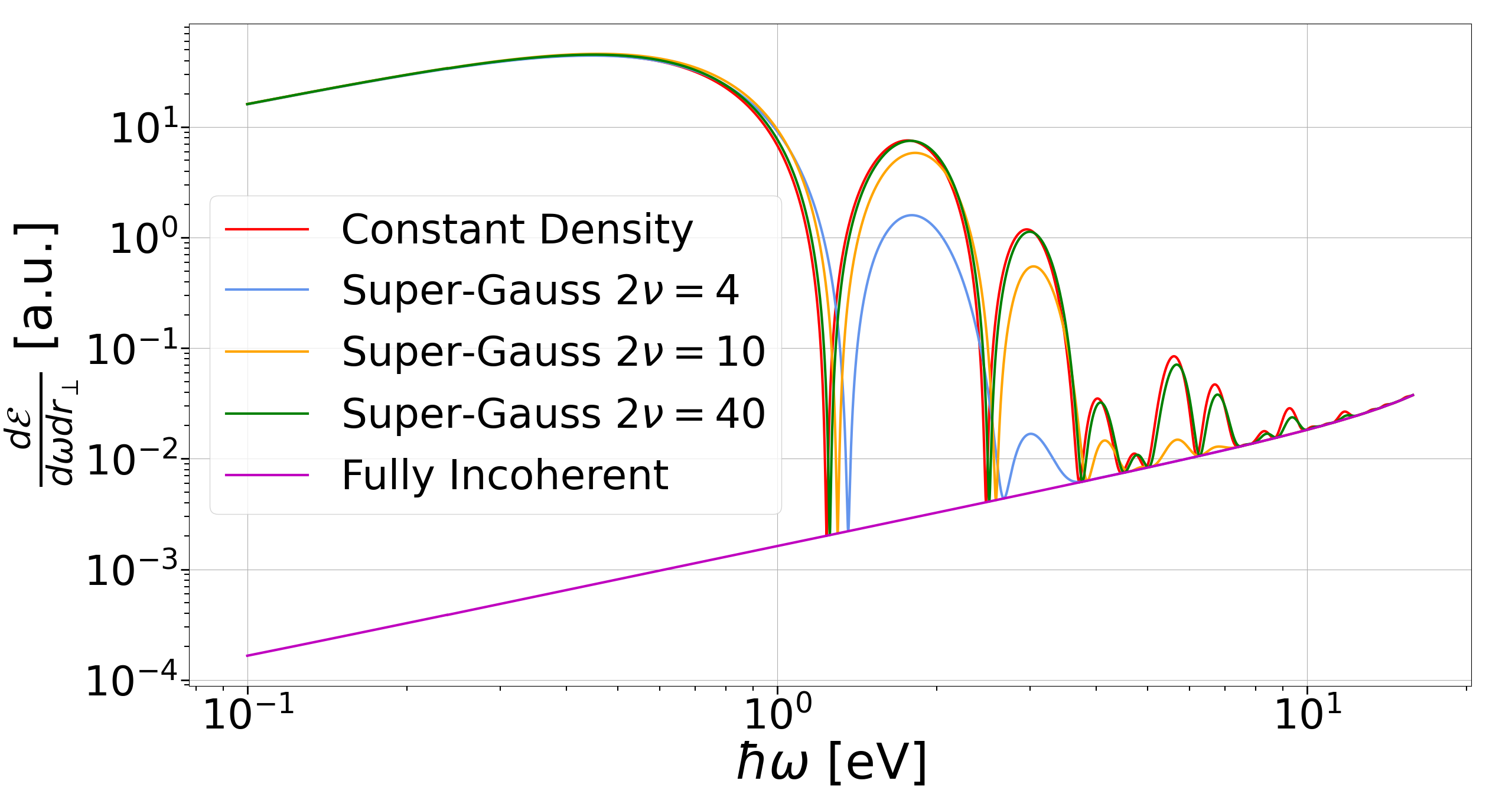}
\caption{Cherenkov energy spectrum ("top spectrum") by a cylindric constant-density beam and super-Gaussian beams with exponents $2\nu = 4,10,40$ as defined in eq. \eqref{rho_definitions}, as a function of photon energy. Parameters are $n_0 = 2.687 \cdot 10^{19} \, \text{cm}^{-3}, \,r_{\perp} = 1 \,\text{mm}, \, N_e = 10^5, \,\gamma = 2000, \, R_b = 20 \, \mu\text{m}, \, L_b = 1 \, \mu\text{m}$. "Fully Incoherent" case sets $\vert S\vert^2 = N_e$, whereas "Coherent" case utilizes full expression \eqref{PhaseFactorGeneralFormula}.}
\label{Figure6}
\end{figure}

\indent Let us therefore consider an intermediate case, the super-Gaussian beam as defined in eq. \eqref{rho_definitions}, which provides a transition from the Gaussian distribution to the constant-density distribution when the parameter $\nu$ is increased, since $\varrho_{g_\nu} \approx \varrho_c$ for $\nu \gg 1$. For visual orientation, the one-dimensional super-Gaussian distribution $\mathcal{N}_\nu \exp\left(-x^{2\nu}\right)$ is shown in fig. \ref{Figure5} for some values of the exponent $2\nu$.\\
\indent As figs. \ref{Figure5} and \ref{Figure6} demonstrate, the rectangular limit of the super-Gaussian distribution for high $\nu$ coincides with the constant-density cylinder case. The difference in the maximum coherently enhanced frequency for the various $\nu$ values is clearly visible, showing the dependence of the range of coherent enhancement on the electron beam shape. For the lowest super-Gaussian with exponent $2\nu = 4$, the coherent enhancement extends up to about $\hbar\omega\approx 4\,\text{eV}$, for $2\nu = 10$ up to about $\hbar\omega\approx 7\,\text{eV}$ and finally for $2\nu=40$ the distribution is essentially equivalent to the cylindric constant-density beam case, with coherent enhancement reaching up to about $\hbar\omega \approx 10\,\text{eV}$.

\indent We do not give expressions for the maximum coherently enhanced frequency for the super-Gaussian distribution since they have no closed analytical form and therefore are of limited practical use. Physically motivated and also looking towards fig. \ref{Figure6}, we can conclude that the maximum coherently enhanced frequency $\omega_{\text{max}}$ for a super-Gaussian distribution with the corresponding beam parameters will always lie in between the ones for the Gaussian distribution and the constant-density cylinder, i.e.

\begin{align*}
\omega_{\text{max},g} < \omega_{\text{max},g_\nu} < \omega_{\text{max},c}\,,
\end{align*}

\noindent with $\omega_{\text{max},g_\nu} \approx \omega_{\text{max},c}$ for $\nu \gg 1$.

\indent In summary, we have demonstrated that the coherent Cherenkov spectrum is sensitive to the beam shape through the change in maximum coherently enhanced frequencies and the shape of the emission spectrum at higher frequencies, which can enable one to extract information about the beam geometry through the spectrum.\\
\indent "Smooth" beam densities have less range of coherent enhancement in $\omega$ as compared to beam densities with steep ramps, which was shown through the super-Gaussian beam density with increasing exponent $2\nu$. Furthermore, beam shapes with plateaus like the super-Gaussian or cylindric constant-density beam, show oscillations in the spectra corresponding to effective plateau lengths. Comparing with figs. \ref{Figure5} and \ref{Figure6}, we see that with increasing $\nu$ the effective plateau lengths of the super-Gaussian distribution increase and as a result, the oscillation periods corresponding to longitudinal and transverse coherence in the spectrum decrease.

\begin{subsection}{Coherent Cherenkov Photon Number Spectrum}
\indent Finally, we shortly discuss the photon number spectrum

\begin{align*}
\frac{dN_{\gamma}}{d\hbar\omega} = \frac{1}{\hbar^2\omega}\frac{d\mathcal{E}}{d\omega}\,.
\end{align*}

\begin{figure}[H]
\includegraphics[width=8.8cm]{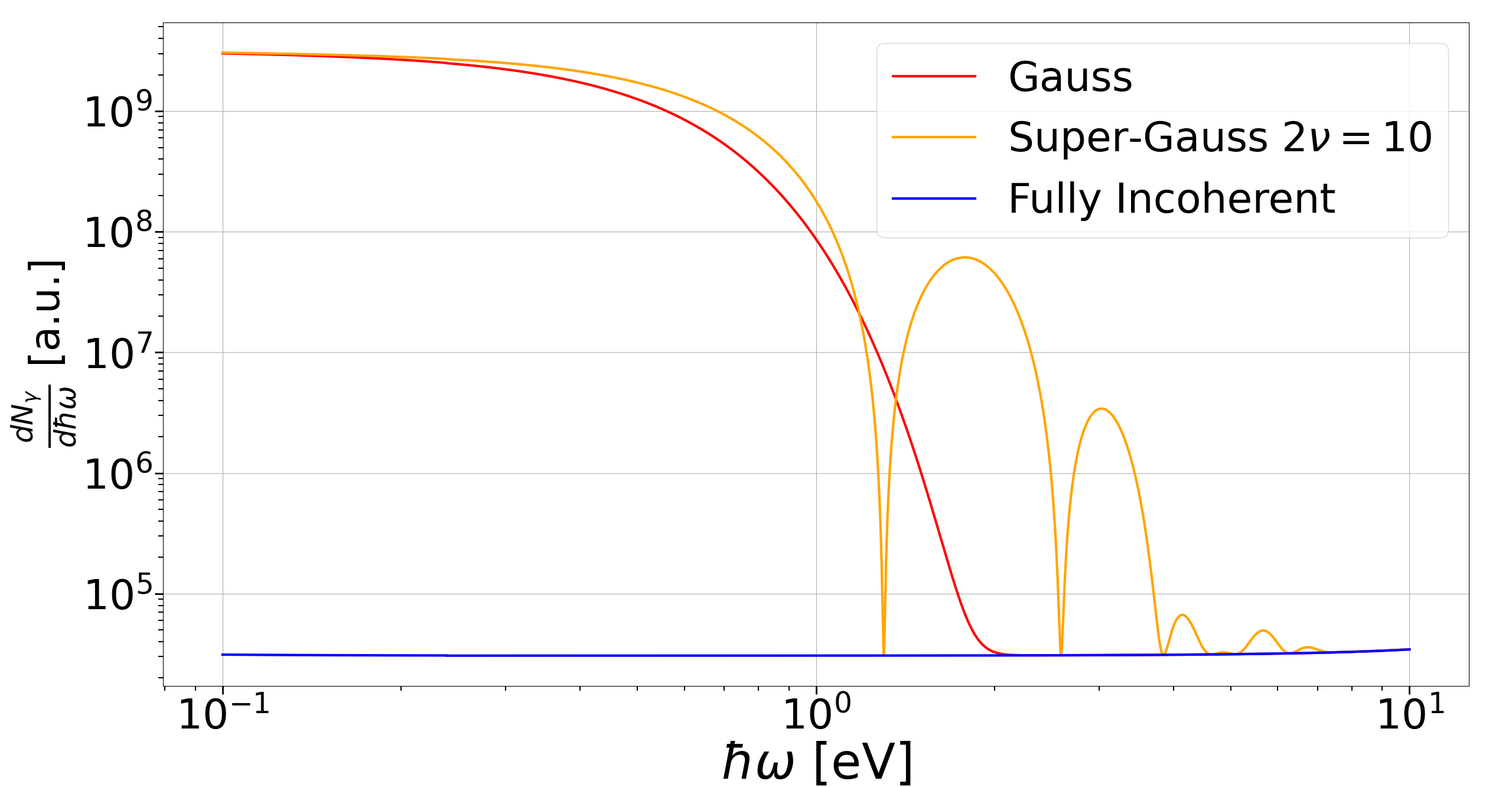}
\caption{Photon number spectrum of photons arriving on ring-shaped detector with inner and outer radii $R_1 = 1\,\text{mm},\,R_2 = 2\,\text{mm}$, as a function of photon energy. The results are shown for a Gaussian beam and super-Gaussian beam with $2\nu=10$ as defined in eq. \eqref{rho_definitions}. Parameters are $n_0 = 2.687 \cdot 10^{19} \, \text{cm}^{-3},\, N_e = 10^5, \,\gamma = 2000, \, R_b = 20 \, \mu\text{m}, \, L_b = 1 \, \mu\text{m}$. "Fully Incoherent" case sets $\vert S\vert^2 = N_e$, whereas "Coherent" case utilizes full expression \eqref{PhaseFactorGeneralFormula}.}
\label{Figure7}
\end{figure}

\noindent Taking $\gamma=2000$ again, the photons are essentially only emitted close to the forward direction as we saw before, thus we assume a detector as shown in fig. \ref{Figure1} positioned around the point at which the beam exits the medium. Integrating the top spectrum \eqref{dE_domega_top_result2} over $r_{\perp}$ from the inner radius $R_1 = 1 \, \text{mm}$ to the outer radius $R_2 = 2 \, \text{mm}$ yields

\begin{align*}
\frac{dN_\gamma}{d\hbar\omega} \approx \frac{e^2}{\hbar^2 c^2} \, \frac{\sqrt{\frac{\varepsilon v^2}{c^2} - 1}}{\frac{\varepsilon v^2}{c^2}} \left(R_2 - R_1\right) \vert S\vert^2\,.
\end{align*}

\noindent The only dependence on $\omega$ is contained in $\varepsilon(\omega)$, which changes only slightly towards the right end of the $\omega$-interval chosen in this case. Hence the coherent photon spectrum is essentially just the coherence factor multiplied by the constant single-electron spectrum.\\
\indent The photon spectrum is shown in figure \ref{Figure7}, with a Gaussian and a super-Gaussian beam shape with $2\nu=10$, using the same set of parameters as were in previous figures.

\end{subsection}

\begin{subsection}{Other Sources of Radiation}
\indent Finally let us discuss competing effects, where the most relevant for emitted radiation in this case are transition radiation and bremsstrahlung.\\
\indent If we assume our medium to be much bigger than a cylinder with radius $R_f$ and length $L_f$, see. eqs. \eqref{AsympRadius} and \eqref{AsympLength}, transition radiation is negligible compared to Cherenkov radiation \cite{TransitionRadiation}. Since this assumption was indeed a condition for use of the theory presented in this paper, we can thus neglect transition radiation.\\
\indent Concerning bremsstrahlung, we can make some estimates for the parameters used in the previous plots. The geometrical radiation length of helium for $n_0 = 2.687 \cdot 10^{19} \, \text{cm}^{-3}$ is \cite{Tsai}

\begin{align*}
L_{\text{rad}} \approx 5.3 \, \text{km}\,.
\end{align*}

\indent Taking a much smaller medium length, we can use the approximate thin-target expression for the bremsstrahlung spectrum \cite{Tsai}

\begin{align*}
\frac{d\mathcal{E}}{d\hbar\omega}\Biggl\vert_{\text{Br}} = \frac{L_{\text{med}}}{L_{\text{rad}}} \left( \frac{4}{3} - \frac{4\hbar\omega}{3\gamma m_e c^2} + \frac{\hbar^2\omega^2}{\gamma^2 m_e^2 c^4} \right)\,,
\end{align*}

\noindent where $L_{\text{med}}$ denotes the length of the helium gas target. Since the photon energies $\hbar\omega$ we are interested in are much smalller than the beam-electron energy $\gamma m_e c^2$, the last two terms in the bracket are negligible.\\
\indent The bremsstrahlung spectrum of the electron beam with $L_{\text{med}} \sim 30 \, \text{cm}$ and $N_e = 10^5$ is then -- if we assume full coherence to set an upper limit on the intensity of bremsstrahlung -- approximately constant with the value

\begin{align*}
N_e^2\frac{d\mathcal{E}}{d\hbar\omega}\Biggl\vert_{\text{Br}} \approx 7.5 \cdot 10^5\,.
\end{align*}

\noindent Taking the results for the Cherenkov radiation from fig. \ref{Figure3} and integrating them from the inner radius $R_1 = 1\,\text{mm}$ to the outer radius $R_2 = 2\,\text{mm}$ of a ring-shaped detector (as in fig. \ref{Figure7}) positioned around the point where the beam exits the medium, we can estimate the minimum coherent Cherenkov radiation energy spectrum in the considered frequency range as

\begin{align*}
\frac{d\mathcal{E}}{d\hbar\omega}\Biggl\vert_{\text{Ch}}\gtrsim 10^5\,.
\end{align*}

\indent Since we set an outer radius of $R_2 = 2\,\text{mm}$, according to the arguments made at the end of section \ref{SectionTheory}, this assumes a medium length extending around the point on the beam propagation axis at distance $R_2/\vartheta \approx 23\,\text{cm}$ from the detector by at least a few formation lengths $L_f$. Since $L_f \lesssim 2.5\,\text{cm}$ for the interval $0.1\,\text{eV}\leq \hbar\omega\lesssim 20\,\text{eV}$ which we chose for our numerical results, the medium length $L_{\text{med}} = 30\,\text{cm}$ should be sufficient for this comparison.\\
\indent Consequently, even fully coherently enhanced bremsstrahlung would give a contribution on approximately the same scale as the minimum of the coherent Cherenkov spectrum.\\
\indent Moreover, a crucial difference between these two types of radiation is their angular distribution. For such low frequencies compared to the beam-electron energies $\sim 1 \, \text{GeV}$ as we considered in our numerical results, the dependence of the bremsstrahlung emission on the emission angle $\theta$ is of the form \cite{Jackson}

\begin{align*}
\frac{\left(1 + \gamma^4 \theta^4\right)}{\left(1 + \gamma^2 \theta^2\right)^4}\,.
\end{align*}

\noindent This distribution sharply peaks at $\theta = 0$ with width $\sim 1/\gamma$. Since the Cherenkov angle for the parameters used before was $\vartheta \approx 1/116$, the width $\sim 1/2000$ of the bremsstrahlung spectrum will be much smaller which enables good angular separation between the Cherenkov spectrum and the bremsstrahlung spectrum. Consequently, higher beam-electron energy is beneficial for the separation of the bremsstrahlung and Cherenkov spectra. To be specific, for the parameters chosen here, there will be practically no bremsstrahlung arriving on the proposed ring detector as long as the medium length $L_{\text{med}}$ is far below $2$ meters.
\end{subsection}
\section{Conclusion}\label{SectionConclusion}
We have considered the theory of coherent Cherenkov radiation by relativistic electron beams for for (i) a Gaussian, (ii) a cylindric constant-density and (iii) a super-Gaussian beam. We demonstrated that the coherent enhancement in the emission of Cherenkov radiation extends to photon frequencies for which the longitudinal and transverse coherence lengths $\lambda_\parallel = \frac{\lambdabar}{\cos(\vartheta)}$ and $\lambda_\perp = \frac{\lambdabar}{\sin(\vartheta)}$, respectively still significantly exceed the respective inter-electron distances in the beam, where $\lambdabar$ is the reduced photon wavelength and $\vartheta$ the Cherenkov angle.\\
\indent The coherence effects most easily arise at low photon frequencies. The maximum coherently enhanced frequency can be increased by raising the transverse coherence length, which can be achieved by increasing electron energy and choosing a medium with refractive index closer to 1. Thus we applied this theory to an ultra-compact and highly relavistic electron beam penetrating a helium gas, with a beam length of few femtoseconds, beam radius $20\,\mu\text{m}$ and beam-electron energy $1\,\text{GeV}$. We demonstrated coherent enhancement up to the optical regime for a Gaussian beam shape. We also illustrated that the range of coherent enhancement in $\omega$ as well as the shape of the coherent Cherenkov spectrum in the region of $\omega$ where coherence vanishes, depends strongly on the beam shape.\\
\indent In our considerations, the beam energy was taken as highly relativistic and the size of the medium along the beam propagation was assumed to be much larger than the formation length $L_f$. Under such conditions, transition radiation is negligible. Moreover the bremsstrahlung and Cherenkov contributions in the photon emission essentially do not overlap due to a much sharper angular distribution of the bremsstrahlung emission in the longitudinal direction.\\
\indent Consequently, for electron beams such as considered in section \ref{SectionResults}, Cherenkov radiation would not be masked by competing effects like bremsstrahlung or transition radiation and the maximum coherently enhanced frequency as well as the shape of the spectrum could be observed in the optical or low-ultraviolet frequency regime to gain information about the beam shape.
\section*{Acknowledgment}
Sami Kim gratefully acknowledges funding by the Studienstiftung des deutschen Volkes.
\section{Appendix}\label{SectionAppendix}
Starting from eq. \eqref{Maxwell}, we apply the Fourier transform in the following form:

\begin{align*}
\phi({\bf r}, t) &= \frac{1}{(2\pi)^2} \int_{-\infty}^{+\infty}d\omega\int_{\mathbb{R}^3} d^3{\bf k} \, \tilde{\phi}({\bf k}, \omega) \, e^{i({\bf k} \cdot {\bf r} - \omega t)}\,,\\
{\bf A}({\bf r}, t) &= \frac{1}{(2\pi)^2} \int_{-\infty}^{+\infty}d\omega\int_{\mathbb{R}^3}  d^3{\bf k} \, \tilde{{\bf A}}({\bf k}, \omega) \, e^{i({\bf k} \cdot {\bf r} - \omega t)}\,,\\
\tilde{\rho}({\bf k}, \omega) &= \frac{e}{2\pi} \delta(\omega - vk_z) \sum_{\ell=1}^{N_e} e^{i{\bf k}_{\perp}\cdot{\bf r}_{\perp,\ell}} \, e^{-i\omega t_\ell}\,,\\
\tilde{{\bf j}}({\bf k}, \omega) &= v\tilde{\rho}({\bf k}, \omega) \, {\bf e}_z\,.
\end{align*}

\noindent We then obtain the solutions for the transformed potentials in Fourier space:

\begin{align*}
\tilde{\phi}({\bf k}, \omega) &= \frac{2e}{\varepsilon} \frac{\delta(\omega - vk_z)}{k^2 - \frac{\varepsilon\omega^2}{c^2}}\sum_{\ell=1}^{N_e} e^{-i{\bf k}_{\perp}\cdot{\bf r}_{\perp,\ell}} \, e^{i\omega t_\ell}\,,\\
\tilde{{\bf A}}({\bf k}, \omega) &= \frac{\varepsilon v}{c} \tilde{\phi}({\bf k}, \omega) \, {\bf e}_z\,.
\end{align*}

\noindent Here we formally applied the $\hat{\varepsilon}$-operator to the potentials in Fourier space which transforms it into a complex scalar function evaluated at the frequency $\omega$. For the fields, we write

\begin{align*}
\tilde{{\bf E}}({\bf k}, \omega) &= -i{\bf k}\tilde{\phi}( {\bf k}, \omega) + \frac{i\omega}{c} \tilde{{\bf A}}({\bf k}, \omega)\,,\\
\tilde{{\bf B}}({\bf k}, \omega) &= i{\bf k} \times \tilde{{\bf A}}({\bf k}, \omega)\,.
\end{align*}

\noindent Inserting the expressions for the potentials and integrating over $d^3{\bf k}$ gives eq. \eqref{E_omega} for ${\bf E}_\omega$.\\
\indent The $z$-component of ${\bf E}_{\omega}$ can be expressed as

\begin{align*}
E_{\omega, z} &= \frac{2ie}{(2\pi)^{2/3}c^2} \, \omega e^{i\frac{\omega z}{v}} \Bigl(1 - \frac{c^2}{\varepsilon v^2}\Bigr) \sum_{\ell=1}^{N_e} e^{i\omega t_\ell} \, I_\ell\,,
\end{align*}

\noindent with $I_\ell = \int_0^{2\pi}d\varphi_k\int_0^{\infty} dk_{\perp} k_{\perp}\frac{e^{ik_{\perp}\Delta r_{\perp,\ell}\cos(\varphi_k + \Delta\varphi_\ell)}}{k_{\perp}^2 - k_0^2}$ where $\Delta\varphi_\ell$ denotes some $\ell$-dependent constant phase.\\
\indent Using $e^{iz\sin(\vartheta)} = \sum_{n=-\infty}^{\infty} J_n(z)e^{in\vartheta}$ \cite{AbrStegun}, where $J_n(z)$ is the Bessel function of the first kind, we continue with the integral separately:

\begin{align*}
I_\ell &= 2\pi \int_0^{\infty}dk_{\perp} \frac{k_{\perp}J_0(k_{\perp}\Delta r_{\perp,\ell})}{k_{\perp}^2 - k_0^2}\,.
\end{align*}

\noindent The Bessel functions of the first and second kind of order 0 have the following integral representations \cite{AbrStegun}:

\begin{align*}
J_0(x) &= \frac{2}{\pi} \int_1^{\infty} dt\, \frac{\sin(xt)}{\sqrt{t^2-1}}\,,\\
Y_0(x) &= -\frac{2}{\pi} \int_1^{\infty} dt\, \frac{\cos(xt)}{\sqrt{t^2-1}}\,.
\end{align*}

\noindent Using the representation for $J_0$, it follows

\begin{align*}
I_\ell = \frac{4\pi}{\pi} \int_1^{\infty} dt \, \frac{1}{\sqrt{t^2-1}} \int_0^{\infty} dk_{\perp} \, \frac{k_{\perp} \sin(k_{\perp}\Delta r_{\perp,\ell}t)}{k_{\perp}^2 - k_0^2}\,.
\end{align*}

\indent The integrand in the integral over $dk_{\perp}$ is now an even function in $k_{\perp}$, so we can extend the lower boundary to $-\infty$ and evaluate it using the Residue Theorem. Then inserting both Bessel function representations again, we finally obtain

\begin{align*}
I_\ell &= i\pi^2 H_0^{(1)}(k_0 \, \Delta r_{\perp,\ell})\,,
\end{align*}

\noindent where $H_{\nu}^{(1)}(z)$ is the Hankel function of the first kind of order $\nu$. Inserting the result for $I_n$ in the expression above, we finally arrive at the expression \eqref{E_omega_z_Result} for $E_{\omega, z}$.\\
\indent The transverse component of eq. \eqref{E_omega} can be written as

\begin{align*}
{\bf E}_{\omega,\perp} &= -\frac{2ie}{(2\pi)^{3/2}v\varepsilon} e^{i\frac{\omega z}{v}} \sum_{\ell=1}^{N_e} e^{i\omega t_\ell} (-i{\bf \nabla}_{{\bf r}_{\perp}} I_\ell) \,.
\end{align*}

\noindent Since $\frac{d}{dz}H_0^{(1)}(z) = -H_1^{(1)}(z)$ \cite{AbrStegun}, we arrive at eq. \eqref{E_omega_perp_Result} for ${\bf E}_{\omega,\perp}$.\\
\indent Continuing from eq. \eqref{E_omega_approx_result}, we can use eq. \eqref{FourierFields} to write

\begin{align*}
\text{Re}\Bigl[{\bf E}_{\omega} \times {\bf B}_{\omega}^* \Bigr] &\,= \text{Re}\left[\begin{pmatrix}
E_{\omega,x} \\ E_{\omega,y} \\ E_{\omega,z}
\end{pmatrix} \times \begin{pmatrix}
B_{\omega,x}^* \\ B_{\omega,y}^* \\ B_{\omega,z}^*
\end{pmatrix} \right]\\
&= \text{Re}\left[ \frac{\varepsilon^* v}{c} \begin{pmatrix}
-E_{\omega,z} \, E_{\omega,x}^* \\ -E_{\omega,z} \, E_{\omega,y}^* \\ \vert E_{\omega,\perp}\vert^2
\end{pmatrix} \right]\,.
\end{align*}

\noindent Inserting the definition eq. \eqref{dE_domega_mantle_help}, we thereby arrive at eq. \eqref{dE_domega_mantle_result1} for the mantle spectrum.\\
\indent Similarly for the top spectrum, we can use eq. \eqref{FourierFields} and the definition \eqref{dE_domega_top_help} such that

\begin{align*}
\frac{d\mathcal{E}}{d\omega} &= \frac{c}{2\pi}\int_0^{2\pi} d\varphi \int_{R_1}^{R_2} dr_{\perp} \, r_{\perp} \text{Re}\left[\frac{\varepsilon^* v}{c} \vert E_{\omega,\perp}\vert^2 \right]\\
&\approx \frac{e^2}{2\pi c^2} \, \omega \, \xi_2(\omega) \int_0^{2\pi} d\varphi \int_{R_1}^{R_2} dr_{\perp} \, e^{-2\text{Im}(k_0)r_{\perp}} \vert S\vert^2\\
&= \frac{e^2}{c^2} \, \omega \, \xi_2(\omega) \frac{e^{-2\text{Im}(k_0)R_1} - e^{-2\text{Im}(k_0)R_2}}{2\text{Im}(k_0)} \, \vert S\vert^2\,,
\end{align*}

\noindent which is equivalent to eq. \eqref{dE_domega_top_result1.5}.\\
\indent Assuming the existence of some fundamental period $T$ of the oscillating product $\sin^2(ax)\cos^2\left(bx - \frac{3\pi}{4}\right)$, it follows

\begin{align*}
\frac{1}{T}\int_0^T \,\sin^2(ax)\cos^2\left(bx - \frac{3\pi}{4}\right) dx = \frac{1}{4}\,.
\end{align*}

\noindent In principle, such a fundamental period $T$ only exactly exists if $\frac{a}{b}$ is rational, but we can take an approximate period otherwise, still justifying the result for use in our physical argument for eq. \eqref{CoherenceCondition_Cylinder_Case1} since the oscillation amplitude falls off quickly.\\
\indent In section \ref{SubsectionCoherenceFactor_AngularRelations}, the super-Gaussian density case is normalized through

\begin{align*}
N_e &\overset{!}{=} 2\pi\mathcal{N}_\nu \int_{-\infty}^{\infty} d\eta \, e^{-\left(\frac{\eta}{L_b/2}\right)^{2\nu}} \int_0^{\infty} dr_{\perp} \, r_{\perp} e^{-\left(\frac{r_{\perp}}{R_b}\right)^{2\nu}}\\
&= \frac{2\pi R_b^2 L_b \, \mathcal{N}_\nu}{(2\nu)^2} \int_0^{\infty} d\eta' \eta'^{\frac{1}{2\nu} - 1} e^{-\eta'} \int_0^{\infty} dr_{\perp}' \, r_{\perp}'^{\frac{1}{\nu} - 1} e^{-r_{\perp}'}\\
&= \frac{\pi R_b^2 L_b \, \Gamma\left(\frac{1}{2\nu}\right)\Gamma\left(\frac{1}{\nu}\right)}{2\nu^2} \, \mathcal{N}_\nu\,.
\end{align*}

\noindent The phase factor $S$ in the coherence limit \eqref{PhaseFactor_IntegralCondition} for the super-Gaussian beam density involves the product of two integrals

\begin{align*}
&\quad\int_0^{\infty}d\zeta \, \cos\left(\alpha \zeta\right) e^{-\zeta^{2\nu}} \int_0^{\infty} d\xi\, \xi J_0(\beta \, \xi) e^{-\xi^{2\nu}}\,.\\
\end{align*}

\noindent Generally, we can write

\begin{align*}
\int_0^{\infty} dx \, x^\mu e^{-x^{2\nu}} &= \frac{1}{2\nu} \int_0^{\infty} dx' \, x'^{\frac{\mu + 1 - 2\nu}{2\nu}} e^{-x'} = \frac{\Gamma\left(\frac{\mu + 1}{2\nu}\right)}{2\nu}\,.
\end{align*}

\noindent Inserting Taylor expansions for the cosine and the Bessel function in the original integrals and utilzing the integral above, we can carry out the integration of each part of the sum, giving a corresponding Gamma function.\\
\indent Therefore, the phase factor $S$ for the super-Gaussian distribution is given by

\begin{align*}
S_{g_\nu} &= N_e \, \left[ \sum_{\ell_1 = 0}^{\infty} \frac{(-1)^{\ell_1} (\frac{\omega L_b}{2v})^{2\ell_1}}{(2\ell_1)!} \frac{\Gamma\left(\frac{2\ell_1 + 1}{2\nu}\right)}{\Gamma\left(\frac{1}{2\nu}\right)} \right]\\
&\quad\;\times\left[ \sum_{\ell_2 = 0}^{\infty} \frac{(-1)^{\ell_2} \left(\frac{\omega R_b}{2v} \sqrt{\frac{\varepsilon v^2}{c^2} - 1}\right)^{2\ell_2}}{(\ell_2!)^2} \frac{\Gamma\left(\frac{\ell_2 + 1}{\nu}\right)}{\Gamma\left(\frac{1}{\nu}\right)} \right]\,.
\end{align*}

\noindent For the Gaussian and constant-density beams, we have

\begin{align*}
S_g &= N_e \, \exp\left(-\frac{1}{4}\frac{\omega^2}{v^2}\left(\left[\frac{\varepsilon v^2}{c^2} - 1\right] R_b^2 + \left(\frac{L}{2}\right)^2\right)\right)\,,\\
S_c &= N_e \, \frac{\sin\left(\frac{\omega}{v}\frac{L_b}{2}\right)}{\frac{\omega}{v}\frac{L_b}{2}} \, \frac{J_1\left(\frac{\omega R_b}{v} \sqrt{\frac{\varepsilon v^2}{c^2} - 1}\right)}{\frac{1}{2} \frac{\omega R_b}{v} \sqrt{\frac{\varepsilon v^2}{c^2} - 1}}\,.
\end{align*} 

\noindent Taking the absolute value squared and dividing by $N_e^2$ leads to the expressions for $\Pi$ in eq. \eqref{PhaseFactors_Pi_Gaussian&Constant}.



\begin{thebibliography}{33}
\bibitem{SuperGauss} Super-Gaussian beam shapes are well known for laser pulses; they can be produced experimentally and used for electron acceleration \cite{SuperGauss1,SuperGauss2,SuperGauss3}
\end{thebibliography}
\end{document}